\documentclass[pre,aps,12pt]{revtex4}
\usepackage{epsfig}
\begin{document}

\title{DYNAMICAL SYMMETRIES IN NANOPHYSICS}
\author{$^1$K. Kikoin, $^{1,2}$Y. Avishai and $^3$M.N. Kiselev}
\affiliation{$^1$Department of Physics, Ben-Gurion University of the
Negev, Beer-Sheva 84105, Israel\\
$^2$ Ilse Katz Center for Nano-Technology,  Ben-Gurion University
of the Negev, Beer-Sheva 84105, Israel\\
$^3$Institut f\"ur Theoretische Physik, Universit\"at W\"urzburg,
D-97074 W\"urzburg, Germany}
\date{\today}

\begin{abstract}
The main goal of this short review is to demonstrate the relevance
of dynamical symmetry and its breaking to one of the most active
fields in contemporary condensed matter physics, namely, impurity
problems and the Kondo effect. It is intended to expose yet
another facet of the existing deep and profound relations between
quantum field theory and condensed matter physics. At the same
time, the reader should not get the impression that this topics is
limited to abstract entities and that it is governed by
complicated and sophisticated mathematical structure. On the
contrary, it will be shown that many of the concepts introduced
below are experimentally relevant. It is one of the rare occasions
where the parameters of a dynamical group (e.g the number $n$ of a
group $SO(n)$) can be determined experimentally.

We elaborate on the role of dynamical symmetry in a special
subdiscipline of condensed matter physics, which enters under the
name of correlated impurity problems. It is concerned with the
physics which is exposed when the system is composed of strongly
correlated localized electrons on the one hand, and itinerant
electrons on the other hand. Experimentally, it can be realized in
quantum dots and other artificial objects which can be controlled
by external parameters such as magnetic fields and gate voltages.
Recently, impurity problems have been shown to be realizable also
on single molecules. The size of the pertinent impurities then
reduces to nanometers and the relevant scientific activity is
referred to as nano-science. At low temperatures, the most
striking manifestation of impurity problems is the Kondo effect,
which is present in bulk systems and, in the last few years,
appears to be an indispensable ingredient in nano-science.

The relevant Hamiltonian in impurity problem is a generalized
Anderson hamiltonian which, under certain conditions can be
approximated by a generalized spin Hamiltonian encoding the
exchange interactions between localized and itinerant electrons.
If the nano-object is, in some sense, simple, (for example, a
single localized electrons), there is no extra symmetry in the
problem. On the other hand, if the quantum dot is complex in some
sense, (for example, it contains an even number of electrons or it
contains several potential wells) the effective spin Hamiltonian
reveals a dynamical symmetry. In particular it includes, beyond
the standard spin operators, new sets of vector operators which
are the analogues of the Runge-Lenz operator familiar in the
physics of the hydrogen atom. These operators induce transitions
between different spin multiplets and generate dynamical symmetry
groups (usually $SO(n)$) which are not exposed within the bare
Anderson Hamiltonian. Like in quantum field theory, the most
dramatic aspects of dynamical symmetry in the present context is
not its relation with the spectrum but, rather, the manner in
which it is broken.

We will review the role of dynamical symmetry and its
manifestation in several systems such as planar dots, vertical
dots, semiconductor nano-clusters, and complex quantum dots, and
show how these dynamical symmetries are broken by exchange
interactions with itinerant electrons in the metallic electrodes.
We will then develop the concept within numerous physical
situations, explaining the concept of dynamical symmetry in Kondo
co-tunnelling, including the interesting case of Kondo
co-tunnelling in an external magnetic field. As it turn out, the
concept of dynamical symmetry is meaningful also in systems out of
equilibrium, and the case of dynamical symmetries at finite
frequencies will also be addressed.
\end{abstract}
\maketitle

\section{Dynamical and hidden symmetry in quantum mechanics}

The concept of symmetry in quantum mechanics has had its golden
age in the middle of the last century. In that period the beauty,
elegance and efficiency of group theoretical physics, from
classification of hadron multiplets, isospin in nuclear reactions,
 the orbital symmetry in Rydberg atoms,
 point-groups in crystallography,
translational symmetry in solid state physics, and so on. At the
focus of all these studies is the symmetry group of the underlying
Hamiltonian. Using the powerful formalism of symmetry operators
and their irreducible representations,  the energy spectrum of an
Hamiltonian with a given symmetry could be found in an elegant and
economical way. Exploiting the properties of discrete and
infinitesimal rotation and translation operators, general
statements about the basic properties of quantum mechanical
systems could be formulated in a form of theorems (Bloch theorem,
Goldstone theorem, Adler principle, etc). The intimate relation
between group theory and quantum mechanics has been exposed in
numerous excellent handbooks (see, e.g. \cite{Hammer}).

A somewhat more subtle aspect featuring group theory and quantum
mechanics emerged and was formulated later on, that is, the
concept of dynamical symmetry. The notion of dynamical symmetry
group is distinct from that of the familiar symmetry group. To
understand this distinction in an heuristic way let us recall that
all generators of the symmetry group of the Hamiltonian ${\cal H}$
encode certain integrals of the motion, which commute with
${\cal H}$. These operators induce all transformations which
conserve the symmetry of the Hamiltonian, and may have
non-diagonal matrix elements only within a given irreducible
representation space of ${\cal H}$. On the other hand, dynamical
symmetry of ${\cal H}$ is realized by transformations implementing
transitions between states belonging to {\it different}
irreducible representations of the symmetry group. One may then
say that the irreducible representation of the dynamical group of
an Hamiltonian requires consideration of the energy spectrum as a
whole (see \cite{Mama} for discussion of various aspects of
dynamical symmetries in quantum mechanics).

A special case of dynamical symmetry in quantum mechanics is that
of a {\it hidden} symmetry, where additional degeneracy exists due
to an implicit symmetry of the interaction. In particular, such
hidden symmetry is a characteristic feature of Coulomb interaction
in the hydrogen atom and the isotropic oscillator \cite{Eng}. This
form of hidden symmetry (implemented by the Coulomb interaction)
results from an explicit invariance of $1/r$ potentials in a four
dimensional energy/momentum space \cite{Fock}, and manifests
itself in the degeneracy of states with different orbital angular
momentum $l$.

A mathematically compact and elegant way to describe various kinds
of symmetry operations {\sl in many-particle quantum problems} is
by means of the so called configuration change operators
introduced by J. Hubbard in the context of his seminal work on the
"Hubbard model" \cite{Hub}. These so called Hubbard operators are
constructed, in fact, in terms of Dirac's ket/bra operations in
the space of eigenvectors $|\Psi\rangle$ of the Hamiltonian ${\cal
H}$. In other words, let $E_\Lambda$ be an eigenstate of the
Hamiltonian, that is, $({\cal H}-E_\Lambda)|\Lambda\rangle=0.$
Then a configuration change operator is defined as
\begin{equation}
X^{\Lambda\Lambda'}=|\Lambda\rangle\langle\Lambda'|.
\end{equation}
Following the accepted terminology, we refer to these operators
as {\it Hubbard operators}.
 In the original Hubbard model these
operators were intended to describe excitations in the narrow
electron bands under condition that the on-site Coulomb repulsion
exceeds the kinetic energy of electron tunnelling between
neighboring sites. Then the Hubbard operators are termed as {\it
Bose-like} if the pairs of states $(\Lambda,\Lambda')$ belong to
the same charge sector of the Fock space $\{\Lambda\}$ or if the
charge of the system $N$ is changed by an even number $\Delta N$
as a result of configuration changes $(\Lambda' \to\Lambda)$. In
case of an odd $\Delta N$ the Hubbard operator is termed as {\it
Fermi-like}. Note that the  one-site part ${\cal H}_0$ of the
Hubbard Hamiltonian and the Dirac's projection operator ${\cal P}=
|\Psi\rangle\langle\Psi|$ can be rewritten in terms of diagonal
Hubbard operators,
\begin{eqnarray}
{\cal H}_0& = & \sum_{\Lambda}E_{\Lambda} |\Lambda\rangle\langle
\Lambda| \equiv\sum_{\Lambda}E_{\Lambda} X^{\Lambda \Lambda} ~, \label{HX} \\
{\cal P}& = &  \sum_{\Lambda}X^{\Lambda \Lambda}.  \label{PX}
\end{eqnarray}
The algebra of Hubbard operators is determined by
the commutation relations
\begin{equation}\label{xcom}
[X^{\Lambda_1\Lambda_2},X^{\Lambda_3\Lambda_4}]_\mp=X^{\Lambda_1\Lambda_4}
\delta_{\Lambda_2\Lambda_3} \mp X^{\Lambda_3\Lambda_2}
\delta_{\Lambda_1\Lambda_4}
\end{equation}
and the sum rule
\begin{equation}
\sum_\Lambda X^{\Lambda\Lambda}=1.
\end{equation}
 Here the sign $-$ or $+$ is chosen for
"Bose-like" and "Fermi-like" operators respectively. If at least
one operator in the brackets is Bose-like, the minus sign
prevails. Spin flips and electron-hole excitations are examples of
Bose-like configuration changes, and addition or removal of
electron to the system implies a Fermi-like configuration change.
The  algebra  of Hubbard operators is predetermined by the
right-hand side of Eq. (\ref{xcom}). If the commutation relations
form a closed set of equations, one may speak of an algebra, which
reflects the dynamical symmetry of the manifold $\{\Lambda\}$.

To characterize the dynamical symmetry of a quantum  mechanical
object in terms of Hubbard operators, we consider a system
described by a Hamiltonian ${\cal H}_0$ whose eigenstates
$|\Lambda\rangle=|M\mu\rangle$ form a basis for an irreducible
representation of some Lie group $\sf G_0$. It should be
emphasized that the interaction is included in ${\cal H}_0$. Using
the commutation rules (\ref{xcom}), one easily finds that
\begin{equation}\label{comm}
[X^{\Lambda\Lambda^{\prime}},{\cal H}_0] =
-(E_{M}-E_{M^{\prime}})X^{\Lambda \Lambda^{\prime}}.
\end{equation}
for the Hamiltonian (\ref{HX}). The Hubbard operators $X^{\Lambda
\Lambda^{\prime}}$, which describe transitions between states
belonging to the same irreducible representation with
$E_{M}=E_{M^{\prime}}$, commute with the Hamiltonian and,
according to the general theory of quantum mechanical symmetries,
they form  generators of the symmetry group of the Hamiltonian
${\cal H}_0$. If the states $\Lambda$ and $\Lambda^{\prime}$
belong to different irreducible representations of the group $\sf
G_0$, the corresponding operators $\widetilde X^{\Lambda
\Lambda^{\prime}}$ do not commute with ${\cal H}_0$. The whole set
of operators $X^{\Lambda \Lambda^{\prime}}$ and $\widetilde
X^{\Lambda \Lambda^{\prime}}$ form a closed algebra, provided the
Hamiltonian ${\cal H}_0$ possesses the definite dynamical symmetry
group ${\sf D}_0$. An interesting case (hidden symmetry) emerges
if the interaction part of ${\cal H}_0$ results an additional
degeneracy of states belonging to different irreducible
representations $M$ and $M'$: Certain linear
combinations of Hubbard operators, which
describe the corresponding transitions form
a set of vector operators
characterizing the hidden symmetry of the system. This
scenario of emergence of new operators due to the special
form of the interaction
is analogous to the appearance of the Runge-Lenz vector operator, which
describes the hidden $SO(4)$ symmetry of
the manifold of bound states (closed orbits) in an attractive
Coulomb potential. These operators do not appear explicitly in
the Hamiltonian ${\cal H}_0$, but rather, enter the Casimir
relation, which describe the symmetry invariants. Another
manifestation of dynamical symmetry arises when an accidental
degeneracy of eigenstates occurs, which does not reflect the
symmetry of the Hamiltonian.

Leaving more detailed discussion of the origin of hidden and
dynamical symmetries for the last chapter containing the
mathematical Addendum, we briefly describe here  several examples
of Hamiltonians which exhibit definite dynamical symmetry. A
simple nontrivial Hamiltonian whose associated dynamical symmetry
is characterized by a non-Abelian Lie group is that governing the
physics of a pair of electron spins coupled by an exchange
interaction. This Hamiltonian describes, for example
 the spin spectrum of
hydrogen molecule, as well as spin dimers which are constituents
of various complex molecules, spin ladders, etc. The pertinent
spectrum of spin excitations consists of a singlet $E_S$ and a
triplet $E_T$. The energy gap between these states is due to
exchange interaction $J=\Delta E_{TS}=E_T-E_S$, which can either
be positive (antiferromagnetic coupling) or negative
(ferromagnetic coupling). The dynamical symmetry of the $\{S,T\}$
manifold is that of the $SO(4)$ group. Two vectors generating this
group are constructed by means of Hubbard operators (\ref{comm})
in the  following way:
\begin{eqnarray}
S^+ & = & \sqrt{2}\left(X^{10}+X^{0-1}\right),~ S_z =
X^{11}-X^{-1-1}.
 \label{m.1} \\
R^+ & = & \sqrt{2}\left(X^{1S}-X^{S-1}\right), ~R_z =
-\left(X^{0S}+X^{S0}\right). \nonumber
\end{eqnarray}
Here ${\bf S}$ is the spin 1 operator, while the second vector
${\bf R}$ describes $S/T$ transitions. Its appearance is due to
dynamical symmetry of a {\it spin rotator}. Below we call
operators of this type $R$-operators
The corresponding Lie algebra
$o_4$ is exhausted by the commutation relations
\begin{eqnarray}
&&[S_\alpha,S_\beta]  =
ie_{\alpha\beta\gamma}S_\gamma,~[R_\alpha,R_\beta]=
ie_{\alpha\beta\gamma}S_\gamma,~
[R_\alpha,S_\beta]=ie_{\alpha\beta\gamma}R_\gamma . \label{comm1}
\end{eqnarray}
Here ($\alpha,\beta,\gamma$ are Cartesian coordinate indices, and
$e_{\alpha\beta\gamma}$ is the anti-symmetric Levi-Civita tensor).
These two vector operators are orthogonal, ${\bf S\cdot R} = 0,$
and the representation is fixed by the Casimir operator is ${\bf
S}^2+ {\bf R}^2 =3.$ In terms of these operators the Hamiltonian
${\cal H}_0$ (\ref{HX}) with $\Lambda=S,T\mu,$ and $\mu=0,\pm1$
acquires the form
\begin{equation}
{\cal H}_0=\frac{1}{2}\left(E_T {\bf S}^2 + E_S {\bf R}^2\right)+
Const. \label{1.3a}
\end{equation}
In this simple Hamiltonian all states belong to the same spin
sector of the Fock space, $N=1, \Delta N=0$.

The second example is again concerned with an elementary object,
the Wannier-Mott exciton. It is characterized by transitions with
$\Delta N=2$. Here the manifold $\{\Lambda\}$ consists of the
ground state   (completely occupied valence band and no
excitations, $\Lambda=G$ with $N=0$) and two excitonic states
(bound electron-hole pair in a singlet and triplet states,
$\Lambda=S,T\mu$ with $N=2$ ). The system of Hubbard operators
describing two singlets, one triplet and all allowed transitions
between these states generate (via linear combinations) the group
$SO(5)$. Here, beside the vector $\bf S$ describing the triplet
exciton, there are two $R$-vectors ${\bf R_1, R_2 }$ and a scalar
operator $A$ describing G/T, S/T and G/S transitions. Altogether,
there are ten operators whose linear combination generate the
group $SO(5)$. Explicitly, these generators are expressed in terms
of Hubbard operators as follows:
\begin{eqnarray}
&& R_1^+  = \sqrt{2}\left(X^{1G}-X^{G1}\right), ~\
R_{1z}  =  -\left(X^{0G}+X^{G0}\right),\nonumber \\
&& R_2^+  = \sqrt{2}\left(X^{1S}-X^{S1}\right), ~\
R_{2z}  =  -\left(X^{0S}+X^{S0}\right),\nonumber \\
&& A= i(X^{GS}-X^{SG}). \label{SE}
\end{eqnarray}
To close the algebra the commutation relations (\ref{comm1}) which
are valid for $R_{l\alpha}$ ($l=1,2$) should be completed by
\begin{eqnarray}
&& [R_{l\alpha},R_{1\beta}]=i\delta_{\alpha\beta}A,
\label{comm2} \\
&& [A,R_{l\alpha}]=iR_{1\alpha},\; \;\;[A,S_{l\alpha}]=0.
\nonumber
\end{eqnarray}
The system of commutation relations (\ref{comm1}), (\ref{comm2})
is that of the $o_5$ algebra, and the manifold $\{G,S,T\}$ obeys
an $SO(5)$ dynamical symmetry. The Casimir operator determining
the pertinent representation of the $SO(5)$ group in this case is
${\bf S}^2 + {\bf R}^2 + {\bf R_1}^2 +A^2=4$. In terms of these
operators the exciton Hamiltonian ${\cal H}_0$ acquires the form
\begin{equation}
{\cal H}_0=\frac{1}{2}\left(E_{G} {\bf R}_1^2 + E_T {\bf S}^2 +
E_S {\bf R}^2 \right)+ Const. \label{1.3b}
\end{equation}

Our last example of a relatively simple Hamiltonian possessing
dynamical symmetry is the $s$-shell of a hydrogen atom (or
hydrogen-like impurity in semiconductor), with a neutral state
H$^0$ and ionized states H$^\pm$ included in the manifold of
eigenstates. Now the index $\Lambda$ acquires four values,
$\Lambda=0,\sigma,2$. Here $\Lambda=0$ stands for an empty s-shell
(positive ion H$^+$), $\Lambda=\sigma=\uparrow,\downarrow$
corresponds to the neutral state H$^0$ occupied by an electron
with spin $\sigma$, and $\Lambda=2$ means double electron
occupation (negative ion H$^+$). In this system sixteen operators
inducing transitions within a given charge sector (that is,
$X^{\sigma\sigma'}, X^{00}, X^{22}$) as well as those with $\Delta
N=\pm 1$ (namely, $X^{\sigma 0}, X^{\sigma 2}, X^{0\sigma},
X^{2\sigma})$ and $\Delta N=\pm 2$ (explicitly $X^{20}, X^{02}$)
are involved in the set of generators of the dynamical group
$SU(4)$. This generic model is interesting {\it per se}, but it
also plays an important role in the physics of strongly correlated
electron systems. Indeed, the Hamiltonian ${\cal H}_0$ (\ref{HX})
with $\Lambda=0,\sigma,2$ describes the "elementary cell" of the
famous Hubbard Hamiltonian \cite{Hub}, and its symmetry properties
are the key ingredient for studying the structure of its
excitation spectrum.

In many generic situations, the state $\Lambda=2$ is quenched by
strong Coulomb repulsion between two electrons occupying the
elementary shell. The dynamical group of this reduced Hamiltonian
is the group $SU(3)$ with eight generators (two vectors, ${\bf
R}_1$, and ${\bf R}_2$ and two scalars $A^+$ and $A^-$):
\begin{eqnarray}
R_1^z=\frac{1}{2}\left(X^{00}-X^{\uparrow\uparrow}\right),~~
R^+_1=X^{0\uparrow}, \nonumber \\
R_2^z=\frac{1}{2}\left(X^{00}-X^{\downarrow\downarrow}\right),
~~R_2^+=X^{0\downarrow} \label{1.4a} \\
A^+=\frac{\sqrt{3}}{2}X^{\uparrow\downarrow},
~~A^-=\frac{\sqrt{3}}{2}X^{\downarrow\uparrow}, \nonumber
\end{eqnarray}
The closed $u_3$ algebra is given by the following commutation
relations:
\begin{eqnarray}
&&[R_i^z,R_i^\pm ]  =  \pm R_i^\pm ,~~~
[R_i^+,R_i^-]=2R_i^z,~~i=1,2
\nonumber \\
&&[R_1^z,A^\pm]=\mp\frac{1}{2}A^\pm,~~~[R_2^z,A^\pm]=\pm\frac{1}{2}A^\pm
\nonumber \\
&& [R_1^+,A^+]=\frac{\sqrt 3}{2}R_2^+,~~[R_1^-,A^-]=-\frac{\sqrt
3}{2}R_2^-, \nonumber \\
&&[R_2^+,A^- ]=\frac{\sqrt 3}{2}R_1^+,~~[R_2^-,A^+]=-\frac{\sqrt
3}{2}R_1^-, \nonumber\\
&&[R_1^-,A^+]=[R_1^+,A^-]=[R_2^-,A^-]=[R_2^+,A^+]=0,\nonumber\\
&& [A^+,A^-]  =  \frac{3}{2}(R_2^z-R_1^z). \label{1.4}
\end{eqnarray}
The Casimir operator is determined as ${\bf R}_1^2+{\bf R}_2^2+
A^+A^-+A^-A^+=3/2$. In terms of above operators the Hamiltonian
(\ref{HX}) with $\Lambda=0,\uparrow,\downarrow,$ acquires the
following form:
\begin{equation}
{\cal H}_0=\frac{4}{3}E_0({\bf R}_1^2+{\bf
R}_2^2)+\frac{4}{3}E_1(A^+A^-+A^-A^+)+ Const.
\end{equation}
Here $E_1=E_\uparrow=E_\downarrow$ (in the absence of an external
magnetic field). A similar $u_3$ algebra can be constructed for
the case where the states $\Lambda=2,\uparrow,\downarrow$ are
included in the reduced manifold. If all four states are included
in the manifold, the dynamical group is $SU(4)$.

In the above examples we formally analyzed the dynamical symmetry
of several simple Hamiltonians without discussing concrete
physical situations where this symmetry may be revealed. One may
find other model systems possessing $SO(n)$ and $SU(n)$ symmetries
(see, e.g., \cite{Flora}), but the above examples illustrate
clearly the general principles of the theory. It is obvious that
transitions described by the operators $\widetilde X$ may be
activated only due to the presence of an external perturbation,
which breaks the symmetry of the Hamiltonian ${\cal H}_0$. Such
perturbation may affect the low-energy part of the spectrum.
Possible physical mechanisms involve, for example, an
 interaction with external reservoirs
(roughly speaking, a reservoir is a thermal bath with continuous
spectrum).

Another example of a symmetry breaking perturbation which
activates a dynamical symmetry is the inclusion of an elementary
cell described by the Hamiltonian ${\cal H}_0$ into a ladder or a
lattice. In both cases the symmetry violation is characterized by
a definite energy scale $\delta\varepsilon$, and only states
separated by an energy gaps $\Delta \leq \delta\varepsilon$ from
the ground state are involved in the dynamical processes. External
magnetic field may influence the dynamics of the system by
introducing an accidental symmetry in the spectrum of ${\cal
H}_0$. The high-energy part of the spectrum may be activated by a
time-dependent external field, both electric and magnetic. If this
field is characterized by a frequency $\Omega$, then those states
with $\Delta \sim \hbar \Omega$ are involved.

It is important to emphasize that a quantum mechanical object with
rich enough energy spectrum manifests different types of dynamical
symmetry in different experiments. For example, transitions
between neighboring charge sectors may be involved in spin-related
phenomena only as virtual processes. Electron hopping from site to
site in a ladder or a lattice involves transitions with $\Delta
N=1$ and, maybe spin flips accompanying these transitions.
Resonance excitation excites only inter-level transitions with
definite energy difference, etc. From the group-theoretical point
of view, this "dynamical ambivalence" follows from the fact that
the dynamical groups $SO(n)$ and $SU(n)$ with $n>2$ may be
presented as a product of several simple groups, and the
corresponding algebras may be presented as direct sums of several
subalgebras. This presentation is not unique, and the choice of
factorization procedure depends on the type of interaction
breaking the symmetry of the Hamiltonian ${\cal H}_0$. Several
examples of such reduction will be demonstrated below.

Physical systems of nanometer size appear to be excellent
candidates for investigating dynamical symmetry effects. First,
one may select a few-particle nano-object with definite
quantum-mechanical symmetry and an easily calculable energy
spectrum and describe it by the Hamiltonian ${\cal H}_0$. Second,
the interaction of this object with environment may be changed and
controlled by means of various external fields. Third, one may
fabricate artificial nano-objects with non-trivial symmetry
properties and thus, model numerous structures which cannot be
found in natural atoms, molecules or crystals.

The aim of this small review is to analyze the symmetry properties
of {\it quantum dots} with due emphasis on the concept of
dynamical symmetry. Quantum dots are nanometer size objects
confining a few number of electrons, which are  fabricated by
means of advanced technologies from semiconductor materials. Both
shape and size of quantum dot can be controlled and varied. The
quantum dot is usually integrated within an electrical circuit,
and the leads or wires connecting the quantum dot with a voltage
source partially play the role of electron reservoirs or thermal
baths. The number of electrons in a quantum dot may be easily
tuned by a gate voltage, and there is no principal difficulty in
applying an external magnetic field in any desired direction.
Besides, one may fabricate complex quantum dots . A complex
quantum dot consists of several (simple or elementary) dots with
possible electron tunnelling and/or electrostatic interaction
between them. A special example (which we discuss in the next
section)is a periodic array of self-assembled quantum dots.

\section{Nanostructures as artificial atoms and molecules}

Nanosize quantum dots with controllable occupation and
variable configurations are ideal objects for studying numerous
physical manifestations of dynamical symmetry.
Recall that a quantum dot is an
artificial structure, which consists of finite number of electrons
confined within a tiny region of space.
If the electron de Broglie
wavelength ($\sim 13$ nm) exceeds the confinement radius, the
energy spectrum of electrons in the dot is discrete. As a result,
the dot can be treated as a "zero-dimensional artificial atom".

These nano-objects may be fabricated in numerous ways. Nanosize
silicon microcrystallites embedded in a matrix of amorphous SiO$_2$
\cite{Tsu} are examples of quantum dot obtained by means of
three-dimensional quantum confinement. Nanocrystallites preserving
the structure of bulk elemental, III-V and II-VI semiconductors
were synthesized by the methods of colloidal chemistry
\cite{Collo}. Defect-free quantum dots of various shapes can be
grown as islands built in highly strained host semiconductors
\cite{Sas}. In the latter case, quantum dots may form
self-assembled periodic or nearly periodic two-dimensional
structures. Quantum dots may be fabricated also by imposing
confining electrodes on two-dimensional electron gas formed near
the interface of heterostructures \cite{Plan}. Such structures are
widely used in the experimental studies of single electron
tunnelling and related many-body effects. Quantum dot devices may
be prepared in a form of disks or pillars (vertical quantum dots)
\cite{Vert}. In this case they preserve the cylindrical symmetry,
and the electron energy spectrum acquires the two-dimensional
shell structure analogous to the three-dimensional shell structure
of "natural" atoms. Another example of artificial quantum object
is a nanoscale ring \cite{Ring}.

Further technological advance enabled fabrication of more
sophisticated nano-objects such as (for example) double quantum
dots \cite{Dqd}. By this we mean a quantum dot consisting of two
wells coupled by electrostatic and/or tunnelling interaction. In
the same manner that a simple quantum dot is considered as an
artificial atom, a complex structure such as double quantum dot
can be looked upon as an artificial molecule. The closest natural
analog is the hydrogen molecule H$_2$ or the corresponding
molecular ions H$_2^\pm$.  As far as further complexity is
concerned, there is no principal obstacles against fabrication of
composite quantum dots consisting of more than two wells. They are
the artificial analogs of complex natural molecules.
\begin{figure}[ht]
\includegraphics[width=7cm,angle=0]{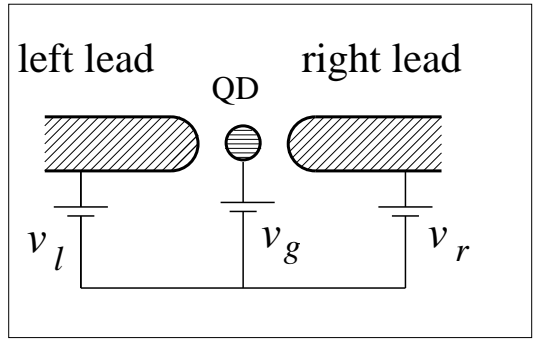}
\caption{Equivalent circuit for a QD connected to two metallic
leads (left and right). Bias voltage is $v_b= v_l-v_r$ , gate
voltage i$v_g$ regulates the number of electrons in the QD.}
\label{f.1}
\end{figure}

The main experimental tool used for studying of electronic
properties of QD is tunnelling current measurements where the
nano-object is incorporated in an electric circuit (Fig.
\ref{f.1}). From the quantum statistical point of view such a
setup can be treated as an artificial atom or molecule in a tunnel
contact with electron reservoirs (metallic leads). The response of
the system to an external bias voltage applied to it is the
tunnelling current. One may learn about the charge and spin
properties of strongly correlated electrons in the QD by means of
measuring its conductance $g$ as a function of temperature,
applied bias, gate voltages and external magnetic field. One may
also study  the optical properties of QD, which are determined by
the linear and non-linear response of QD at high frequencies.  We
are interested in manifestations of dynamical symmetry in these
measurements. The pertinent properties of several types of quantum
dots are discussed below.

\subsection{Planar dots}

\begin{figure}[ht]
\includegraphics[width=9cm,angle=0]{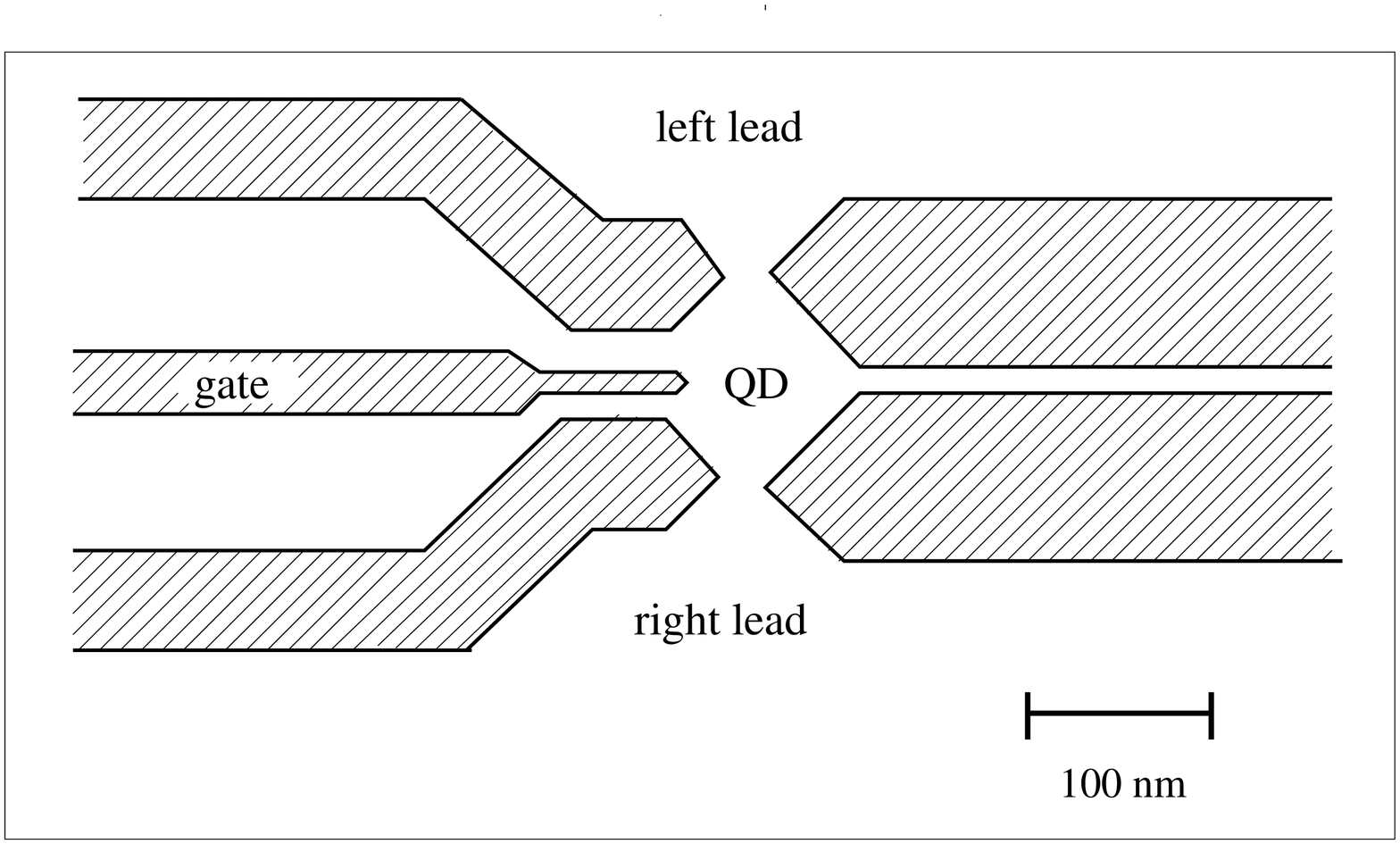}
\caption{Planar quantum dot.  Electrodes imposed on a
semiconductor heterostructure are drawn schematically as shaded
profiles}.\label{f.2}
\end{figure}

A planar (lateral) quantum dot is formed in a 2D depletion layer
on the interface between two semiconductors (usually,
GaAs/Ga$_{1-x}$Al$_x$As). The electrodes imposed on this structure
form both the confining potential well (quantum dot) and the
dot-lead junctions (Fig. \ref{f.2}). If the junction is narrow
enough, a single electron mode connects the dot and the Fermi
reservoirs of electrons in "metallic" leads. The discrete energy
spectrum of electrons in the confined region is characterized  by
 single-particle levels $\epsilon_i$ with typical inter-level
spacing $\delta\varepsilon$ and charging energy $Q$, which is
predetermined by the capacitance $C$ of the dot, namely,
$Q=e^2/2C$.  Typical values of the parameters characterizing
planar quantum dots are $\delta\varepsilon \approx 100-150 \mu
eV$, $Q \approx 500-600 \mu eV$, so the charging energy is large
enough, and it predetermines the character of electron tunnelling
through the dot.

\begin{figure}[ht]
\includegraphics[width=12cm,angle=0]{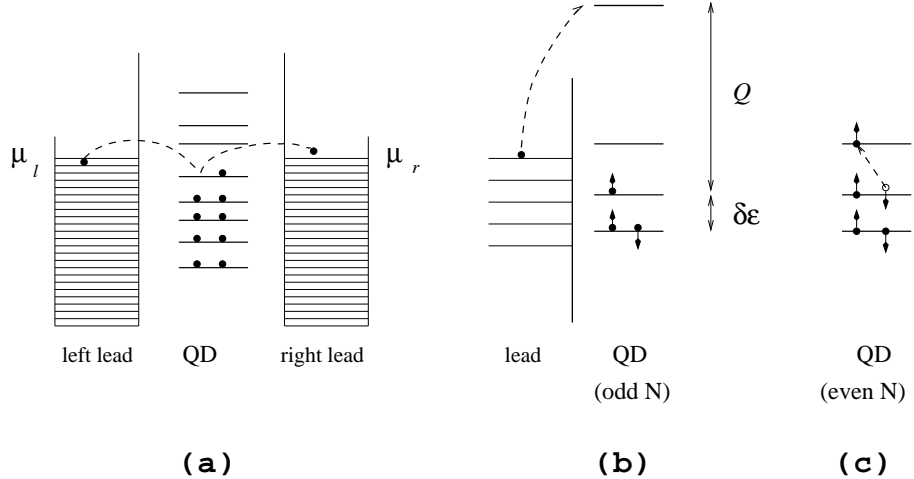}
\caption{Energy levels for electrons in planar quantum dot. (a)
General level scheme for odd ${\cal N}_0$; (b) Zoomed highest
occupied states and lowest unoccupied states for odd ${\cal N}_0$;
(c) Zoomed highest occupied states for even ${\cal N}_0$. Electron
tunnelling processes between leads and QD (a),(b) and within QD
(c) are shown by dashed arrows.}\label{f.3}
\end{figure}

Now one may schematically represent the energy spectrum of a
planar quantum dot coupled by tunnel contacts to the leads  in the
following way: The discrete electron levels in the dot and the
metallic electron continuum in the leads in an equilibrium state
have the same chemical potential $\mu=\mu_L=\mu_R$, where the
indices $L,R$ denote the electron liquid in the left and right
lead, respectively (see  Fig. \ref{f.3}a). However, tunnelling
through planar quantum dots is blocked by the charging energy $Q$
in the weak coupling limit, where the tunnelling rates
$\gamma_{L,R}$ are the smallest energies compared with all other
energy parameters characterizing electrons in the circuit,
\begin{equation}\label{2.1}
\gamma_{L,R}\ll \delta\varepsilon < Q\ll D_{L,R}.
\end{equation}
(here $D_{L,R}$ is the bandwidths characteristic for the kinetic
energy of electrons in the leads). This is the familiar Coulomb
blockade effect responsible for single electron tunnelling in
nano-size devices \cite{Rob}. The Coulomb blockade fixes the
number of electrons $\cal N$ in planar quantum dots. In the
neutral state of planar quantum dots the value of ${\cal N}={\cal
N}_0$ is determined by minimization of the total energy of the dot
${\cal E}({\cal N})$. Then, injection of one more electron into
the initially neutral dot costs an {\it addition energy}
$\Delta{\cal E}_{{\cal N}_0}={\cal E}({\cal N}_0+1)-{\cal E}({\cal
N}_0)-\mu$. This energy equals
\begin{eqnarray}\label{2.2}
\Delta{\cal E}_{{\cal N}_0}=\left\{ \begin{array}{ll} Q-\mu
&{\rm for~odd}~~{\cal N}_0\\
\delta\varepsilon+Q-\mu &{\rm for~even}~~{\cal N}_0
\end{array} \right.
\end{eqnarray}
(see Fig. \ref{f.3}b,c).

The planar quantum dot with odd occupation is characterized by its
spin 1/2, whereas a dot with even occupation is generically in a
singlet spin state due to the Pauli principle. It is clear from
Eqs. (\ref{2.2}), that electron tunnelling through planar quantum
dots is possible only provided the additional energy vanishes. To
achieve this resonance condition, a gate voltage $v_g$ is applied
to the dot, so that resonance tunnelling occurs when $\Delta{\cal
E}-v_g=0$. Then, increasing $v_g$ one may reach the resonance
conditions for $\Delta{\cal E}_{{\cal N}}$ with ${\cal N}={\cal
N}_0+1,~{\cal N}_0+2,~{\cal N}_0+3,$ etc. This is the mechanism of
single electron tunnelling, which was realized in various devices
manifesting Coulomb blockaded features \cite{Grab}.

Returning now to the dynamical symmetry aspect of this phenomenon,
one should consider only the low-energy part of the electron
spectrum in planar quantum dots, which is comparable with the
energy scale characteristic for the interaction with the metallic
reservoirs (electrons in the leads). This scale is determined by
the tunnelling rates $\gamma_{L,R}$ and the energy $k_BT_0$ of
more subtle many-particle effects, which will be specified below
($k_B$ is the Boltzmann constant). As has been shown in the
previous section, dynamical symmetries become essential when
intrinsic or extrinsic level degeneracy exists in a nano-object.
Having this in mind, one can ignore most states of planar quantum
dots except the ground state with its energy ${\cal E}({\cal
N}_0)$ and those states which may enter the resonance under the
influence of external fields (gate voltage, magnetic field, etc).
In case of odd occupation, the only possibility is the Coulomb
resonance determined by the condition $\Delta{\cal E}_{{\cal
N}_0}\approx (\gamma_{L,R},~k_BT_0)$. Then only the following
states,
\begin{eqnarray}\label{2.33}
E_\Lambda({\cal N}_0)=\varepsilon_d-v_g,~~~\Lambda=\sigma \nonumber \\
E_\Lambda'({\cal N}_0+1)=2(\varepsilon_d-v_g)+Q,~~~\Lambda'=S,
\end{eqnarray}
should be retained in the Hamiltonian (\ref{HX}). Here
$\varepsilon_d$ is the last occupied electron level in the dot.
Thus, an accidental degeneracy is possible only between the states
belonging to adjacent charge sectors ${ \cal N}=({\cal N}_0,{\cal
N}_0+1)$.

A richer situation emerges for even occupation. In this case the
Coulomb resonance is also possible, but there arises an additional
possibility of dynamical processes within a given charge sector
${\cal N}={\cal N}_0$. The energy states in this sector are
\begin{eqnarray}
E_\Lambda({\cal N}_0)=2(\varepsilon_d-v_g),~~~\Lambda=S \nonumber \\
E_{\Lambda'}({\cal
N}_0)=\varepsilon_d+\delta\varepsilon-2v_g-J_{d},~~~\Lambda=E_T,\nonumber\\
E_{\Lambda''}({\cal
N}_0)=\varepsilon_d+\delta\varepsilon-2v_g,~~~\Lambda=E_S.
\label{2.44}
\end{eqnarray}
The second and third levels above correspond to triplet and
singlet excitonic states, where one of the electrons moves from
the highest occupied level to the lowest empty level. The
parameter $J_{d}$ is the intra-dot exchange
 interaction responsible for the singlet/triplet level splitting.
The symmetry of this manifold is $SO(5)$ according to the
classification scheme given above. If not all these states are
involved in  the interaction with reservoir, the symmetry is
effectively reduced. Several examples of symmetry reduction will
be discussed below.

\subsection{Vertical dots}

Vertical dots have been fabricated in numerous experiments. The
main feature of such device is that the geometrical symmetry of
the dot (usually cylindrical) occurs also in the lead. This
implies that electrons has, beside spin, an additional quantum
number which is attached to them wherever they are. Electrostatic
confinement potential in vertical quantum dots with cylindrical
symmetry can be approximated by a 2D harmonic potential. Such
potential gives rise to shell structure of discrete electron
states. Like in "natural" atoms, these states are classified as
$1s,~2s,~2p,~3s,~3p,~3d...$ and electronic states are filled
sequentially in accordance with Hund's rule. The occupation
numbers corresponding to filled shells can be regarded as "magic
numbers" $\bar{\cal N}$ with maximum addition energy $\Delta{\cal
E}_{\bar{\cal N}}.$ Indeed, the atom-like character of filling
these artificial atoms with tunable number of electrons in the dot
$0\leq {\cal N} \leq 23$ was observed experimentally \cite{mag}.

Like in the case of real atoms the structure of electron shells is
predetermined by the single-particle quantum numbers and full
orbital and spin momenta of electrons in the last partially filled
shell. Having in mind possible manifestations of dynamical
symmetry, we are especially interested in hidden and accidental
degeneracies characterizing the electron spectrum of vertical dot.
These features may be revealed already for non-interacting
electrons in parabolic potential described by the Fock-Darwin
equation \cite{Darfock}. In this case the electron levels are
characterized by the main quantum number $n$, radial number
$n_r=1,~2,~3...$, $z$-component of angular momentum $m=0,~\pm
1,~\pm 2...$ and spin projection $\sigma=\pm 1/2$. The intrinsic

degeneracy of the spectrum is determined by the condition
\begin{equation}\label{2.5}
n=2n_r + |m| +1,
\end{equation}
whereas the discrete energy levels $\epsilon_n$ are determined by
the law $\epsilon_n = n\hbar\omega_0$, where $\hbar\omega_0$ is
the electrostatic confinement energy. Besides, the electrons in
the vertical dot possess orbital diamagnetism, so that the energy
spectrum in the presence of a magnetic field ${\bf B}$ applied
along the cylindrical axis is split in accordance with the
following equation
\begin{equation}
\label{2.66} \epsilon_n(B)=(2n_r + |m|
+1)\hbar\sqrt{\frac{\omega_0^2}{4}+\omega_c^2}-\frac{m}{2}\hbar\omega_c~,
\end{equation}
where $\omega_c=eB/m^*c$ is the cyclotron frequency (the Zeeman
splitting  is negligibly  small in comparison with diamagnetic
shift in GaAs) . It follows from Eqs. (\ref{2.5}) and (\ref{2.6})
that (i) the magic numbers are $\bar{\cal N}=2,~6,~12,~20,$ etc,
and (ii) that crossing of levels with different $n_r,m$ at some
values of magnetic field is feasible (see Fig. {\ref{f.4}a ).
\begin{figure}[ht]
\includegraphics[width=14cm,angle=0]{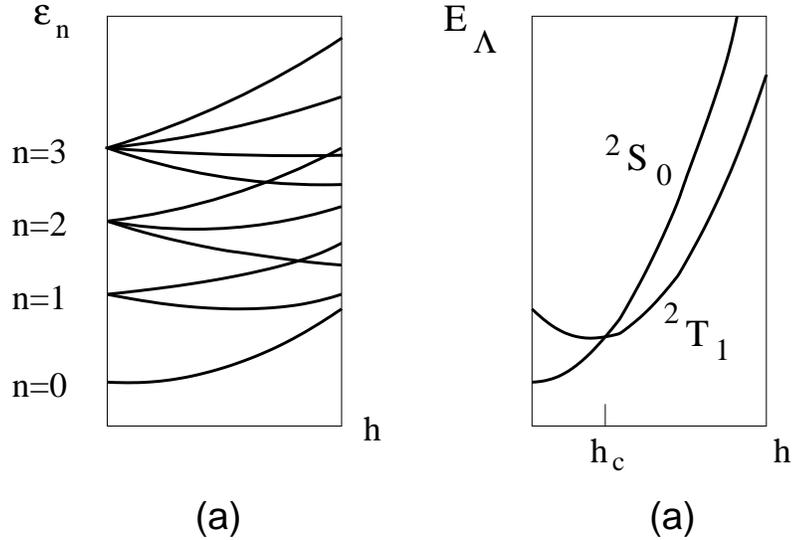}
\caption{Energy levels for electrons in vertical quantum dot. (a)
single electron energy levels [see Eq. (\ref{2.66})]; (b)
two-electron singlet and triplet energy levels.}\label{f.4}
\end{figure}

The level crossing pattern becomes even more complicated when the
electron-electron interaction is taken into account \cite{Chap}.
In the simplest case of $\bar{\cal N}=2$ the two-electron states
are characterized by the total spin $S$ and total orbital
momentum$M$. At zero magnetic field the ground state is singlet
$^2S:~(S=0, M=1)$ and the excited state is triplet $^1T:~(S=1,
M=0)$. These levels cross at certain $h=h_c$ (Fig. {\ref{f.4}b),
and the effective symmetry of the vertical dot is $SO(4)$ in the
vicinity of level crossing. When the second shell is partially
filled (${\cal N}=3,4,5$) more than two multiplets are involved in
the level crossing pattern \cite{Taglia}, so that the dynamical
symmetry becomes really complicated.

\subsection{Semiconductor nanoclusters}
Semiconductor nanocrystals prepared by means of colloid chemistry
methods \cite{Collo} and self-assembled dots grown as islands in
strained lattice-mismatched films \cite{Sas} have regular
geometrical shapes and conserve the bulk crystal structure. As a
result, the discrete electron spectrum in these nanoclusters
retains many feature of Kane electron-hole spectrum of bulk III-V
or II-VI semiconductors \cite{Zung}: the characteristic energy is
in the eV scale, the same
 as in bulk materials. The nomenclature of
discrete highest occupied states in the valence band and lowest
unoccupied states in the conduction band is inherited from the
$sp$-hybrid states in the vicinity of the $\Gamma$ point of the 3D
Brillouin zone.

The excitonic spectrum of these nanoclusters is of primary
interest. Single exciton lines due to interband excitations are
observed in optical and tunnel spectra \cite{Millo}.
High-excitation spectroscopy methods allow experimental
observation of exciton complexes  \cite{array} and even exciton
droplets \cite{Raimond} in arrays of self-assembled dots. These
multiexciton levels may be degenerate due to geometrical and
dynamical symmetries.

The idea of hidden symmetry of excitonic states in quantum dots
\cite{havr2} is borrowed from the theory of excitons in 2D
electron gas in an Integer Quantum Hall Effect regime \cite{Loza}.
The algebra of exciton operators is predetermined by the special
properties of Coulomb interaction for this two-component Fermi
liquid. These operators form a vector ${\bf P}$ with the following
spherical components:
\begin{equation}
P^+=\sum_i e^{\dagger}_{p\sigma}h^{\dagger}_{p,-\sigma},~
P^-=\sum_p h_{p,-\sigma}e_{p\sigma},~
P_z=\frac{1}{2}\left((N^e_\sigma+N^h_\sigma)-N_{tot}\right).
\end{equation}
Here $e^{\dagger}_{p\sigma},~h^{\dagger}_{p,-\sigma}$ are,
respectively, creation operators for the electron and hole with
respective single-particle energies $E_p^{(e,h)}$. The operators
$P^\pm$ create/annihilate an exciton, and the operator $P_z$
measure the population inversion on a shell with orbital
degeneracy $N_{tot}$. In terms of dynamical symmetry, these
operators describe transitions from the ground state (with zero
energy) to a single exciton state. In accordance with Eq.
(\ref{comm}) these operators do not commute with the electron-hole
Hamiltonian ${\cal H}_{eh}$, which includes the single particle
term and the Coulomb interaction $\langle pq|V|rs\rangle$, where
$\{pqrs\}$ are various hole and electron states on a given shell
of the QD. Quite unexpectedly, the commutation relation
(\ref{comm}) reads in this case
\begin{equation}
[P^+, {\cal H}_{eh}]=-E_{0}{\cal H}_{ex},
\end{equation}
where $E_0=E_p^{(e)}+E_p^{(h)}$ is the energy of an electron-hole
pair. All Coulomb contributions cancel each other exactly. This
cancellation is a consequence of hidden symmetry of Coulomb
interaction in quantized two-component Fermi system: interactions
between particles of the same kind coincide with each other and
(with reversed sign) with that of different kinds. It follows from
this property that creation of $n$ excitons of the same kind
described by the operator $(P^+)^n$ costs an energy $nE_0$. This
enables a strong degeneracy of the multi-exciton spectra: for
example, the energy level of biexciton is twice the energy of a
single exciton, etc. Of course, any small perturbation removes the
degeneracy between a singlet-singlet and triplet-triplet states of
a biexciton, but the dynamical symmetry still exists.
\subsection{Complex quantum dots}

The isolated quantum dots considered above are typical examples of
artificial atoms. A double valley quantum dot with weak capacitive
and/or tunnelling coupling between its two wells may be considered
as a simplest case of artificial molecule. Such double quantum
dots (DQD) were fabricated several years ago \cite{Hoff95}. Two
wells in DQD may be identical or have different size; the DQD may
be integrated within an electric circuit either in series or in
parallel; different gate voltages may be applied to each well.
Moreover, one of the two wells may be disconnected from the leads
(side geometry). All these configurations are presented in Fig.
\ref{f.5}. Vertical quantum dots also may form a DQD. Such a
variety of configurations promises even more possibilities for
manifestations of dynamical symmetry effects in artificial
molecules than in artificial atoms.
\begin{figure}[ht]
\includegraphics[width=15cm,angle=0]{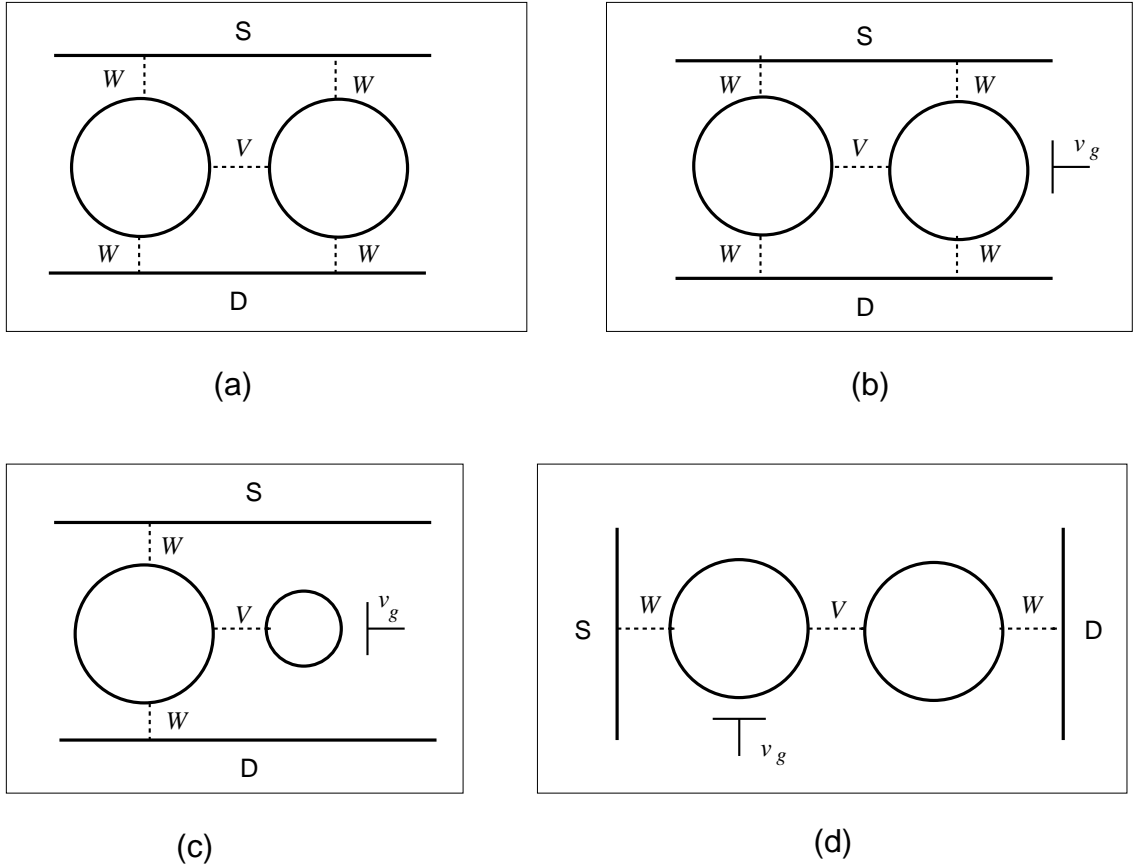}
\caption{Various types of double quantum dots (DQD). Tunnelling
channels are marked by dashed lines, $V$ and $W$ are tunnelling
constants. (a) Symmetric DQD in parallel geometry; (b) biased
symmetric DQD in parallel geometry; (c) asymmetric DQD in side
(T-shaped) geometry; (d) DQD in series geometry.} \label{f.5}
\end{figure}

Let us consider the electron spectra of DQDs shown in Fig.
\ref{f.5} \cite{KA01}. In Section I we discussed dynamical
symmetries of hydrogen-like systems (H$^0$, H$^\pm$ states of
atomic hydrogen). It is clear that the symmetric DQD shown in Fig.
\ref{f.5}a has the main features of hydrogen molecule H$_2$ and
molecular ions H$_2^{\pm}$ for the occupation number ${\cal
N}=2,1,3$, respectively. To exploit this analogy, we assume that
the neutral state of DQD corresponds to ${\cal N}=2$, and each
valley is also neutral when occupied by a single electron, ${\cal
N}_i=1, i=l,r$. Then two valleys are coupled only by tunnel
channel, and in case of strong Coulomb blockade in each valley,
$Q\gg V$ ($V$ is a tunnelling constant), the system is an analog
of H$_2$ in the Heitler-London limit. If the two valleys are
completely identical, the spectrum $E_\Lambda$ consists of two
low-lying spin states $E_{S,T}$ and two charge transfer excitons
$E_{e,o}$ (even singlet and odd triplet), with
\begin{eqnarray}
E_S & = & 2\varepsilon-2V\beta,~~~ E_T  =  2\varepsilon
\label{2.6} \\
E_{o} & = & 2\varepsilon +Q,~~~ E_{e}  =  2\varepsilon +Q +2\beta
V~ . \nonumber
\end{eqnarray}
Here $\varepsilon=\varepsilon_l=\varepsilon_r$ are the
one-electron levels. The spectrum is calculated under the
assumption $\beta=V/Q\ll 1$.  The origin of a narrow spin gap
$\sim \beta V$ and a wide charge transfer gap $\sim Q$ in the
two-electron spectrum is inter-valley tunnelling, which is
possible only in singlet configurations. Fig. \ref{f.6}a
illustrates this mechanism.
\begin{figure}[ht]
\includegraphics[width=12cm,angle=0]{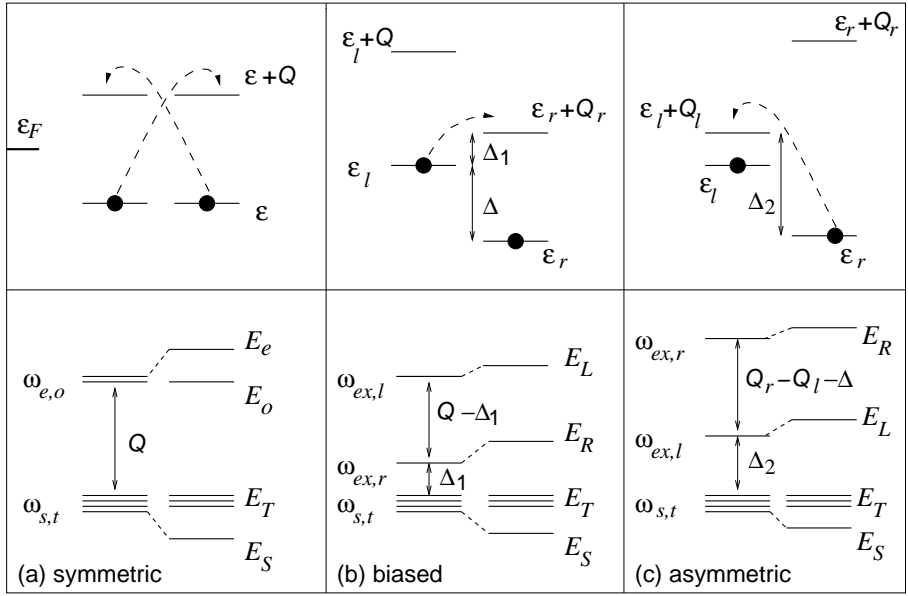}
\caption{Energy level schemes  for symmetric, biased and
asymmetric DQD. Upper panels: single electron levels. Tunnel
processes responsible for formation of charge transfer excitons
are shown by dashed lines. Lower panels: degenerate two-electron
energy levels $\omega_\alpha$ without interwell tunnelling $V$ and
split two-electron levels $E_\Lambda$ renormalized by interwell
tunnelling in second order.}\label{f.6}
\end{figure}

One may break the left-right symmetry by means of an external gate
voltage difference $v_{gl} - v_{gr}$ applied to the DQD
configuration (Fig. \ref{f.5}b). This voltage can be tuned in such
a way that one of two charge transfer excitons becomes "soft"
enough to influence the low-lying spin excitations (Fig.
\ref{f.6}b). In this case, only three levels should be taken into
account when discussing the dynamical symmetry: these are the same
singlet and triplet states plus a singlet charge-transfer exciton
("right" exciton in the geometry illustrated by Fig. \ref{f.6}b.
The corresponding energy levels are
\begin{eqnarray}
E_S & = & \varepsilon_l+\varepsilon_r-2\beta_1 V,~~~ E_T =
\varepsilon_l+\varepsilon_r ,
\nonumber \\
E_R & = & 2\varepsilon_r + Q + 2\beta_1 V~. \label{2.3}
\end{eqnarray}
Finally, the two constituents of a DQD may have different size
(Fig. \ref{f.5}c). We discuss this case in a side geometry, where
the right dot of a smaller  size $R_r$ is disconnected from the
leads.

If $R_r\ll R_l$, then $Q_r \gg Q_l$, and the relevant part of the
spectrum of doubly occupied DQD shown in Fig. \ref{f.6}c is
\begin{eqnarray}
E_S & = & \varepsilon_l+\varepsilon_r-2\beta_2 V, ~~~ E_T =
\varepsilon_l+\varepsilon_r ,
\nonumber\\
E_L & = & 2\varepsilon_l + Q_l + 2\beta_2, \label{2.4}
\end{eqnarray}
In this case the "right"  charge transfer exciton is retained in
the manifold.

It follows from the above analysis, that DQD may manifest $SO(3)$,
$SO(4)$ or $SO(5)$ dynamical symmetries depending on the relevant
physical processes, in which the DQD is involved. The first of
these symmetries is simply a symmetry of a spin $S=1$, while in
the second case, both spin singlet and spin triplet are inveigled.
In the third case, the charge transfer exciton also plays its
part. All these situations will be discussed in subsequent
sections.

Next, the symmetry of a DQD with ${\cal N}=1$ should be exposed.
Again, we consider the case of strong Coulomb blockades in both
valleys, where the doubly occupied states are eliminated. Although
this restriction sounds rather formal in the case of single
electron occupation, it starts to play an essential role when the
interaction with the reservoir is switched on. The dynamical
symmetry of this system was analyzed in \cite{Bord03}. We restrict
ourselves to the case of symmetric or nearly symmetric DQD in a
series geometry (Fig. \ref{f.5}d). If $\varepsilon_l \approx
\varepsilon_r$, then four states have approximately the same
energy, namely left or right dot occupied by an electron with spin
projection $\uparrow$ or $\downarrow$. One may introduce a
pseudospin $T_z=({\cal N}_l-{\cal N}_r)/2=\pm 1/2$, which
describes the electron configuration in the DQD. Denoting the two
components of this pseudospin states as $\pm$, one may construct a
"hyperspin" $\chi_4$ including all four states
\begin{equation}\label{2.7}
\chi_4=\{+\uparrow,+\downarrow,-\uparrow,-\downarrow \}.
\end{equation}
This spinor transforms in accordance with representations of the
$SU(4)$ group. If the four-fold degeneracy of the DQD is violated
by external fields (gate voltage or magnetic field), the remaining
symmetry is $SU(2)$. In particular, one may imagine the situation
when only the pseudospin ${\bf T}$ is involved in the dynamical
symmetry, and spin variable plays part of "flavor" \cite{poh,avi}.
The case of a symmetric DQD with ${\cal N}=3$ may be considered
similarly because of particle-hole and spin up-spin down symmetry
of the nano-object.

Another way to obtain an artificial molecule is to fabricate two
vertically stacked columnar quantum dots coupled by tunnelling
interaction \cite{havr}. Of course, the familiar singlet/triplet
level crossing induced by diamagnetic level shift is expected for
a H$_2$-like molecule with ${\cal N}=2$ \cite{petpar}.
 More complex molecular spectra appear at higher occupations
${\cal N}>2$. An especially exotic state such as a "bipolar"
molecule arises when two vertically stacked quantum dots are
populated by carriers of different kinds -- electrons and holes
\cite{anis}.

Recently it was shown that even richer picture of dynamical
symmetries may arise in triple quantum dots (TQD) with ${\cal
N}=4$ both in parallel and sequential geometries \cite{KKA}. In
addition to the situation of nearly degenerate
singlet-singlet-triplet manifold ($SO(5)$ symmetry) the
triplet-triplet-singlet level crossing can occur in TQD under
certain conditions. The dynamical symmetry of this manifold is
$SO(7)$. Following the arguments of Ref. \cite{Bord03}, it is
clear that the same TQD with ${\cal N}=1$ should possess $SU(6)$
symmetry.

To conclude, the notion of dynamical symmetries is the natural
language, which is extremely useful in the description of quantum
dots. These nano-objects possess a plethora of dynamical
symmetries, which may be unveiled due to interaction with metallic
reservoirs and external fields. Various physical realizations of
these dynamical symmetries will be described in subsequent
sections.
\section{Dynamical symmetries broken by interaction with
reservoir}

As long as the dot is isolated from its environment, it conserves
the symmetry related to its spin state, charge state and as well
as its geometrical symmetry. In a simple case when only spin and
charge are conserved quantities, the generic Hamiltonian of an
isolated QD may be written in terms of these conserved quantities
\begin{equation}\label{3.0}
{\cal H}_{dot}=\sum_{i\sigma} \epsilon_i n_{i\sigma} + Q({\cal
N}-{\cal N}_0)^2-J{\bf S}^2.
\end{equation}
Here the first term describes the discrete spectrum of the QD and
the two other terms stem from electrostatic and exchange
interactions and describe the Coulomb blockade of charge states
(deviating from the neutral state with ${\cal N}={\cal N}_0$) and
the total spin of the dot. A tunnelling contact with reservoir may
break the symmetry of ${\cal H}_{0}$ and, hence, violate spin and
charge conservation. The global spin and charge in a system as a
whole is conserved, but interaction with environment can change
both these quantities in the dot. The simplest example of
interaction, which changes the charge in the dot is electron
tunnelling between the dot and the electronic reservoir, and the
simplest spin-dependent interaction is that of direct exchange
between the dot and the lead electrons.

We discuss first the tunnelling contact. The basic tunnel
Hamiltonian is
\begin{equation}\label{3.1}
{\cal H}_{tun}=\sum_ {\alpha ki\sigma}W_{\alpha k}^{i\sigma}\left(
 c_{\alpha
k\sigma}^\dag d_{i\sigma} + H.c.\right).
\end{equation}
Here $W_{\alpha k}^{i\sigma}$ is the tunnel matrix element,
$d_{i\sigma}$ is the operator destroying the dot electron in the
level i$\epsilon_i$ and $c_{\alpha k\sigma}^\dag$ is the operator
creating an electron in the lead $\alpha$ of metallic reservoir
described by the Hamiltonian
\begin{equation}\label{3.2}
{\cal H}_{band}=\sum_{\alpha=l,r}\sum_{k\sigma}
\varepsilon_{\alpha k\sigma} c_{\alpha k\sigma}^\dag c_{\alpha
k\sigma}.
\end{equation}
Below we discuss the case of weak tunnelling, where the QD is
connected  only by one tunnelling channel with each lead and the
inequalities
\begin{equation}\label{3.3}
W_{\alpha k}^{i\sigma}\ll \delta\epsilon_i\ll Q
\end{equation}
are valid. In this case only the highest occupied levels of QD are
involved in tunnelling, and the energy cost of single electron
tunnelling between the lead and the QD is
$$
E_\pm=\left\{ \begin{array}{ll}
\epsilon_{{\cal N}+1}-\varepsilon_{\alpha k\sigma}, & k<k_F \\
\varepsilon_{\alpha k\sigma}-\epsilon_{\cal N}, & k>k_F
\end{array} \right.
$$

Here $\epsilon_{{\cal N}+1}=E_{\Lambda'}({\cal N}_0+1) -
E_{\Lambda}({\cal N}_0)$ and $\epsilon_{{\cal
N}}=E_{\Lambda}({\cal N}_0) - E_{\Lambda'}({\cal N}_0-1)$ are
electron addition and extraction energies respectively, and $k_F$
is the Fermi momentum of electrons in the lead [cf. Eqs.
(\ref{2.33}) and (\ref{2.44})].

It is convenient to rewrite ${\cal H}_{tun}$ in terms of Hubbard
operators:
\begin{equation}\label{tun1}
{\cal H}_{tun}=\sum_{\alpha
k\sigma}\left[\sum_{\Lambda\lambda}\left(W_{\alpha
k}^{\Lambda\lambda}X^{\Lambda\lambda}c_{\alpha k\sigma} +
H.c.\right)+\sum_{\Lambda\gamma}\left(W_{\alpha
k}^{\gamma\Lambda}X^{\gamma \Lambda}c_{\alpha k\sigma} +
H.c.\right)\right]
\end{equation}
Here the states from the charge sectors ${ \cal N}_0\mp 1$ are
denoted by indices $\lambda$ and $\gamma$, respectively. If the
gate voltages are tuned so that either an addition or extraction
energy is zero, we are at resonance and tunnelling through the dot
is possible. The dot then becomes "transparent" in accordance with
the Breit-Wigner formula for transmission coefficient ${\cal
T}(\varepsilon)$:
\begin{equation}\label{brev}
{\cal T}(\varepsilon)=\frac{\Gamma_l \Gamma_r}{[
\varepsilon-E_{\pm}(k_F)]^2+(\Gamma_l+\Gamma_r)^2/4},
\end{equation}
where $\Gamma_\alpha=\pi\rho_{0\alpha}|W_ \alpha^\pm|^2$ are the
tunnelling rates for the left and right lead, $\rho_{0\alpha}$ is
the density of electron states on the Fermi level of the lead
$\alpha$, $W_ \alpha^\pm$ are the matrix elements for
addition/extraction tunnelling taken from the Hamiltonian
(\ref{tun1}). These processes are responsible for single-electron
tunnelling through quantum dots, which results if the ubiquitous
Coulomb staircase for the current-voltage characteristics
prevails. This form of single electron transistor arises in tunnel
junctions and, as far as the tunnelling current is concerned,
serves as a replacement of the Ohm's law \cite{Grab}.
\begin{figure}[ht]
\includegraphics[width=13cm,angle=0]{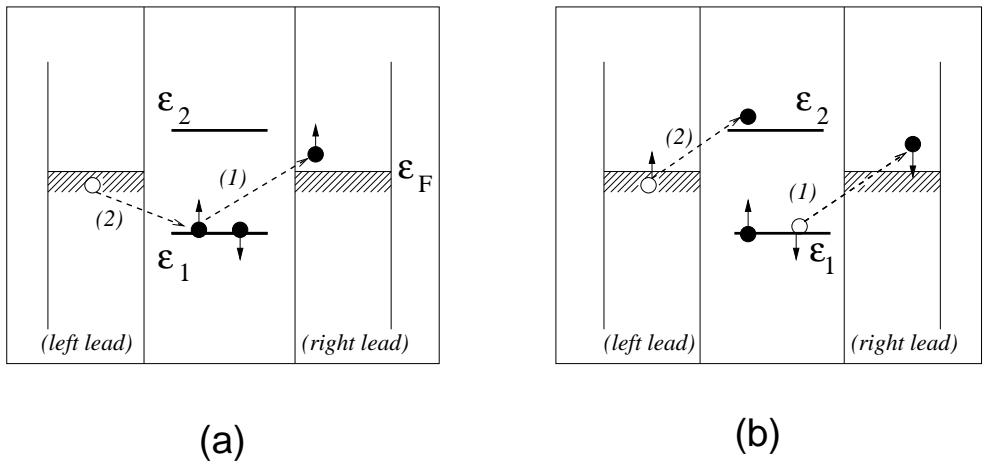}
\caption{Cotunneling through planar QD with odd (a) and even (b)
occupation.}\label{f.7}
\end{figure}

The second order tunnelling processes  described above cannot
reveal any dynamical symmetry of QD because they do not change the
state of the dot (see Fig. \ref{f.7}a). On the other hand,
dynamical symmetries can be exposed by {\it cotunnelling}
processes out of resonance (on the Fermi level) (Fig. \ref{f.7}b):
in the first step of a two-stage cotunnelling process  an electron
from the occupied level $\epsilon_1$ leaves the dot and then, in
the second step, an electron from the same or another lead
replaces it in the empty level $\epsilon_2$. The many-electron
state of a QD changes from $\Lambda_1$ to $\Lambda_2$, and if the
operator $X^{\Lambda_2\Lambda_1}$ belongs to a set of generators
of a dynamical group, this {\it inelastic} cotunnelling may be
described by means of the group theoretical manipulations
introduced above.

The effective vertex $\Gamma$ for this inelastic cotunnelling has
the form
\begin{equation} \label{gamma}
\Gamma_{\alpha
k\sigma,\alpha'k'\sigma'}^{\Lambda_1\Lambda_2}(\varepsilon) \sim
\frac{W_{\alpha } W_{\alpha'}}{(\varepsilon-\delta \epsilon)}
c^{\dagger}_{\alpha k\sigma}c_{\alpha'
k'\sigma'}X^{\Lambda_2\Lambda_1}
\end{equation}
Here $\delta \epsilon$ is the additional energy of the dot in the
process of two-stage cotunnelling.

The fact that cotunnelling process is accompanied by spin
reversal, $\sigma \neq \sigma'$ is crucial. At the bottom line it
means that this two-particle process may be treated also as an
effective exchange, which involves the dot particles and electrons
from both leads. In a situation where the system is set away from
a Coulomb resonance, $E_{\pm}(k_F)\gg W_\alpha $, the only
process, which survives is this indirect exchange (in its full
quantum-mechanical sense), where only the electron residing on the
highest occupied level $\epsilon_h$ of the dot is involved in
cotunnelling. Then $\delta \epsilon =Q$, and the vertex $\Gamma$
combines with direct exchange between the dot and lead electrons
to generate the effective spin Hamiltonian for {\it an elastic}
cotunnelling
\begin{equation}
\label{cotkon} {\cal H}_{cot}=\sum_{\alpha
\alpha'}\sum_{kk'}J_{\alpha\alpha'} {\bf S}\cdot{\bf s}_{\alpha
k,\alpha' k'}.
\end{equation}
with $J_{\alpha\alpha'}=\Gamma_{\alpha
k_F,\alpha'k{^\prime}_F}^{0}(0)$ Here the upper index '0' indicate
that only the spin state of the dot is changed in the course of
cotunnelling.

The Hamiltonian (\ref{cotkon}) has the form of the well known
Kondo Hamiltonian, which was introduced in order to explain the
anomalous shallow minimum in the resistance of
 metals doped be magnetic impurities. This
analogy between magnetic scattering in bulk metal and magnetic
tunnelling through insulating barrier was noticed in the mid 60ies
\cite{App}. Later on, this idea was extended to the physics of
tunnelling through nano-objects such as quantum dots \cite{GR}.
For quite a number of years now, resonance Kondo tunnelling is at
the focus of contemporary physics of quantum dots
 and related objects (such as carbon nano-tubes or other
molecules). There are several reasons for this exceptional
interest. First, the Kondo effect {\it per se} is a complex
many-particle phenomenon, which can be formulated in terms of
integrable 1D problem (using e.g Bethe Ansatz). Second, Kondo
tunnelling manifests itself as an enhanced quasi elastic
resonance, and thus, it can be detected as a zero bias anomaly
(ZBA) in a tunnelling conductance away from Coulomb resonance
peaks in the so called Coulomb windows which form a diamond-shape
lattice of valleys in the plane $(v_b, v_g)$/ \cite{Grab}.
(here $v_{b}$ and $v_{g}$
are external bias and gate voltage, respectively).  These
zero-bias anomalies were detected in planar QD in the late 90ies
\cite{KKK}. Thereafter, similar effects were found in nanotubes
\cite{ext} and vertical dots \cite{Wiel,Sasa}. In the two latter
cases, the anomalies related to the the $SO(4)$ dynamical symmetry
in Kondo tunnelling were noticed for the first time.

\subsection{Dynamical symmetry in Kondo cotunnelling}
{\it Prima facie}, the cotunneling Hamiltonian (\ref{cotkon}) in
conjunction with the Hamiltonian (\ref{3.0}) for the isolated QD
leaves no room for manifestations of dynamical symmetry. The only
operator characterizing the excitations in the QD which are
involved in the interaction with reservoirs is the spin ${\bf S}$.
This exchange interaction, of the form $J{\bf s} \cdot {\bf S}$
breaks the $SU(2)$ symmetry of the QD, and the multiple creation
of electron-hole pairs with spin reversal results in the formation
of a singlet ground state in case when the spin of the dot is
$S=1/2$ \cite{GR}. This situation is realized in QD with odd
occupation number (see Fig. \ref{f.3}b). Indeed in the first
experiments \cite{KKK}, Kondo-type ZBA were observed only in
Coulomb windows with odd ${\cal N}$. Soon after, though,
theoretical predictions about possible realizations of Kondo
mechanism of cotunneling in QD with even ${\cal N}$ in external
magnetic field were reported \cite{PAK,taglia,Eto00}. They were
promptly verified experimentally \cite{ext,Sasa}. Later on,
possible occurrence of Kondo effect in evenly occupied double
quantum dots without magnetic field assistance was suggested
\cite{KA01}.

Most of theoretical investigations pertaining to even ${\cal N}$
appealed implicitly or explicitly to the concept of dynamical
symmetry of the QD, which was discussed in Section II. The
simplest case of even occupation is ${\cal N}=2$ of course. This
situation may be easily realized in planar QD (see Section II.A),
vertical QD (Section II.B) and double QD (Section II.D). Let us
recall that in case of odd ${\cal N}$ where there is a single
electron in the highest occupied level of QD, its spin may be
reversed due to cotunneling together with creation of
electron-hole pair, thus resulting in the effective Hamiltonian
(\ref{cotkon}). A similar process may result in a singlet-triplet
transition in case of two electrons in the highest occupied level.
Thus the vector operator ${\bf R}$ describing these
singlet-triplet transitions (\ref{m.1}) enters the effective
Hamiltonian together with the spin vector operator ${\bf S}$:
\begin{equation}
\label{cotron} {\cal H}_{cot}=\sum_{\alpha
\alpha'}\sum_{kk'}\left(J_{\alpha\alpha'}^{(1)} {\bf S}\cdot{\bf
s}_{\alpha k,\alpha' k'}+ J_{\alpha\alpha'}^{(2)} {\bf R}\cdot{\bf
s}_{\alpha k,\alpha' k'}\right).
\end{equation}

One should remember, however, that in all cases mentioned above
the ground state of the isolated QD is singlet. The energy gap
$\Delta_s=E_T-E_S$ between the triplet and singlet levels is
\begin{eqnarray} \label{gapp}
\Delta_s=\left\{ \begin{array}{ll}
\delta\varepsilon-J_d  & {\rm in~ planar~ QD} \\
J_d & {\rm in~ vertical~ QD} \\
2\beta V & {\rm in ~double~ QD}
\end{array} \right.
\end{eqnarray}
[see Eqs.(\ref{2.44})) and (\ref{2.6})]. The Kondo effect is
characterized by the energy scale $E_K$ (Kondo energy), which is
exponentially small: in the simple case described by the
Hamiltonian (\ref{cotkon}), \begin{equation}\label{EK} E_K =
k_BT_K \sim D_0 \exp(-1/\rho_0J_{max}),
\end{equation}
 where
$J_{max}$ is the maximum of all coupling constants entering this
Hamiltonian and $D_0$ is the effective energy band width for the
electrons in the leads. If $\Delta_s \gg E_K$ Kondo screening is
ineffective and no ZBA in the conductance develops. In accordance
with general principles described in Section I, the dynamical
symmetry enters the physical considerations provided $\Delta_s
\sim E_K$, so the special mechanisms compensating the spin gap
$\Delta_s$ should exist to allow Kondo screening in quantum dots
with even occupation. Such mechanisms exist in all three systems
under discussion. A concrete description of {\it singlet-triplet
crossover} in QDs is given below.

Let us start with vertical dots, where the mechanism of
singlet-triplet crossover is simple from a theoretical point of
view and may be easily realized experimentally. As was mentioned
in Section II.B, the occasional singlet-triplet degeneracy arises
due to diamagnetic shift of single electron levels in an external
magnetic field ( Fig. \ref{f.4}b). To derive the effective spin
Hamiltonian, one should start with the Anderson Hamiltonian
\begin{equation}\label{3.4}
{\cal H}_A={\cal H}_{dot}+ {\cal H}_{band}+ {\cal H}_{tun},
\end{equation}
where the approximation (\ref{3.0}) is {\it not enough} even for
adequate description of the charge sector ${\cal N}=2$, in which
the states  $\{\Lambda=S,T\mu\}$ from the singlet/triplet manifold
should be included. In terms of the group generators this term in
the Anderson Hamiltonian may be rewritten as
\begin{equation}
{\cal H}_{dot}=\frac{1}{2}\left(E_T {\bf S}^2 + E_S {\bf
R}^2\right)+ Q(\hat{\cal N}-2)^2. \label{3.0a}
\end{equation}
Besides, states from adjacent charge sectors, $\{\lambda,\gamma\}$
which are admixed due to the interaction ${\cal H}_{tun}$
(\ref{tun1}) should be taken into account. Projecting out the
virtual transitions to charge states with ${\cal N} = 1,3$ by
means of second order perturbation theory \cite{Hewson} or by
employing the Schrieffer-Wolff (SW) canonical transformation
\cite{SW} (which eliminates ${\cal H}_{tun}$), one maps the
initial problem onto a reduced Fock space with conserved ${\cal
N}=2$. Before performing the SW transformation, it is worthwhile,
in many cases, to execute a rotation in the lead Hamiltonian
\begin{eqnarray}
c_{lk\sigma}&=&c_{1k\sigma}\cos \vartheta_{k\sigma}+
c_{2k\sigma}\sin \vartheta_{k\sigma}\nonumber \\
c_{rk\sigma}&=&-c_{1k\sigma}\sin\vartheta_{k\sigma}+
c_{2k\sigma}\cos\vartheta_{k\sigma}\label{rot}
\end{eqnarray}
with $$\vartheta_{k\sigma}=\arcsin\left(
W_{lk}^\sigma/\sqrt{|W_{lk}^\sigma|^2+|W_{rk}^\sigma|^2}\right)$$
By this transformation the odd combination $c_{2k\sigma}$ is
eliminated from the tunnel Hamiltonian \cite{GR}, and eventually
one arrives at the Hamiltonian (\ref{cotron}) with diagonal
coupling constants
\begin{equation}
\label{cotron2} {\cal H}_{cot}=\sum_{kk'}\left(J^{(1)} {\bf
S}\cdot{\bf s}_{kk'}+ J^{(2)} {\bf R}\cdot{\bf s}_{kk'}\right).
\end{equation}
At this stage, the cotunneling problem can be approached by
standard methods of the theory of Kondo scattering generalized to
the case of $SO(4)$ symmetry, where the "scatterer" is described
by a six-dimensional vector $\{{\bf S,R}\}$ instead of the usual
$SO(3)$ vector ${\bf S}$ for spin 1.

The conventional Kondo problem \cite{Hewson,WT} is easily
manageable at high temperatures $T>T_K$, where perturbation
theory may be constructed by using the dimensionless quantity
$j(\varepsilon)$,
\begin{equation}\label{3.5}
j(\varepsilon)= \rho_0J\ln(D_0/\varepsilon),
\end{equation}
as a small parameter. This approach works until the current energy
parameter $\varepsilon$ is big enough to keep $j(\varepsilon)\ll
1$. The energy $E_K$ defined by the equality $j(E_K)=1$ is defined
as the Kondo energy mentioned above as a characteristic energy of
the Kondo problem. The crossover from the weak Kondo coupling to
the strong coupling regime occurs around $\varepsilon \approx E_K$
[see Eq. (\ref{EK})].

Summation of complicated perturbation series can be carried out by
using the parquet equation technique \cite{Abrik}, but one also
may elegantly pack the same series in the solution of scaling
equations within renormalization group (RG) theory
\cite{Hewson,AM,Zawad,Coza,Andrg} due to occurrence of logarithmic
terms in the effective vertex which indicates that all energy
scales in the Kondo problem are of equal importance. According to
the RG scheme, one investigates the evolution of  vertices and
pertinent propagators by reducing the energy scale $D$ (the
bandwidth of the conduction electrons) in the Hamiltonian ${\cal
H}_{band}+{\cal H}_{cot}$ from the bare bandwidth $D=D_0$ to the
limiting value of $D\sim E_K$. The resulting low-energy spectrum
is universal, which means that there exists a single energy scale
$E_K$ for all thermodynamic and dynamic variables.

If one employs Eq. (\ref{cotkon}) for the cotunneling Hamiltonian,
the scaling flow trajectories are determined by a single equation
for the effective dimensionless exchange vertex ${\cal J}(D)$
\begin{equation}
\label{3.6} d{\cal J}/dL= -{\cal J}^2,
\end{equation}
which should be solved with initial condition ${\cal J}(D_0)=J$.
The solution is then,
\begin{equation}
{\cal J}(D)=\frac{\rho_0J}{1-j(D)},
\end{equation}
which implies that the effective coupling strength flows toward
infinity  when $D\to E_K$ or, in a temperature scale, ${\cal
J}(T)=J/\ln (T/T_K)$. This {\it stable infinite fixed point}
determines the universality class of the Kondo problem. The
scaling flow trajectory is trivial in this case: it is simply a
ray (half line) emanating from the point ${\cal J}(D)=\rho_0J$ and
flows to $\infty$. 
The RG approach cannot describe the
subtle behavior of the system at $T \to 0$, but it encodes generic
feature of Kondo cotunneling, i.e. formation of ZBA in Kondo
regime. In the weak coupling regime $T>T_K$ the formation of the
ZBA in tunnel conductance $g=dI/dV_b,~V_b\to 0$ follows the
scaling form \cite{kng}
\begin{equation}
g(T)/g_0 \sim \ln^{-2} T/T_K,
\end{equation}
where
$$
g_0=\frac{e^2}{h}\frac{4\Gamma_l\Gamma_r}{(\Gamma_l+\Gamma_r)^2},
$$
is the resonance conductance of the quantum dot (see Eq.
\ref{brev}). More sophisticated calculations at $T<T_K$ \cite{Aff}
allow one to describe the appearance of "Abrikosov-Suhl resonance"
on the Fermi level $\varepsilon_F$ responsible for the unitarity
limit for the transmission coefficient ${\cal T}(\varepsilon \to
\varepsilon_F)$.

The situation with the RG flow diagram  for a QD possessing
dynamical symmetry is more complicated. Since  the Hamiltonian
(\ref{cotron2}) contains two effective vertices, one gets a system
of scaling equations \cite{KA01,Eto00}
\begin{eqnarray}
    d{\cal J}_1/dL&= &-[({\cal J}_1)^2+({\cal J}_2)^2],\nonumber\\
    d{\cal J}_2/dL&= &-2{\cal J}_1{\cal J}_2 \label{3.7}.
\end{eqnarray}
The vertex ${\cal J}_2$ describes effective exchange due to S/T
transitions, which are generically inelastic, because one has to
absorb or release the energy of exchange splitting $\Delta_s$ to
excite the singlet state in a situation when the ground state is a
triplet $(\Delta_s<0)$. The scaling procedure for ${\cal J}_2$
stops at $D\sim |\Delta_s|$ and below this energy, only ${\cal
J}_1$ continues to grow with decreasing $D$. Due to this effect
the Kondo tunnelling is no more universal: unlike the conventional
Kondo effect, it is impossible to draw the family of flow scaling
diagrams in the plane $\{{\cal J}_1,{\cal J}_2 \}.$ The flow
trajectory for ${\cal J}_1(D)$ still ends at stable infinite fixed
point, but ${\cal J}_1$ is quenched at an intermediate energy, so
the Kondo temperature becomes a function of $\Delta_s$, which
characterizes this quenching. Since one can change $\Delta_s$ in
vertical QD by varying the magnetic field (see Fig.\ref{f.4}b),
one may reach the point of accidental degeneracy $\Delta_s=0$ and
change its sign. It is easily seen that $T_K$ has maximum at
$\Delta_s=0$. In this case the system (\ref{3.7}) reduces to a
single equation (\ref{3.6}) for the effective vertex ${\cal J}^+ =
{\cal J}_1+{\cal J}_2$. The corresponding Kondo temperature is
\begin{equation}\label{3.9a}
T_{K0}=D_0\exp\left[-1/\rho_0(J^{(1)}+J^{(2)})\right].
\end{equation}
At finite $|\Delta_s|$ the Kondo temperature decreases and at
$|\Delta_s|\gg T_{K0}$ it obeys a kind of universal law
\begin{equation}\label{3.9b}
T_{K}/T_{K0}=(T_{K0}/\Delta_s)^\zeta,
\end{equation}
 where $\zeta\lesssim 1$ is a numerical constant specific for a
 given type of QD \cite{KA01,Eto00,Pust1}.

Unlike the case of vertical dots, the singlet/triplet crossover in
planar dots and double dots with even ${\cal N}$ may be achieved
without the application of an external magnetic field due to
somewhat rich and complicated structure of the low-energy states
in an isolated QD. The most important fact is that the tunnelling
amplitudes $W^{\Lambda\lambda}$ in the Hamiltonian (\ref{tun1})
may be different for singlet ($\Lambda\lambda=S\bar\sigma$) and
triplet ($\Lambda\lambda=T\sigma$) states. In planar QD the origin
of this distinction is that electrons from {\it different} single
electron states $\epsilon_1$ and $\epsilon_2$ are involved in the
formation of $S$ and $T$ states, respectively (see Fig.
\ref{f.7}b). Besides, one should remember that the singlet exciton
also enters the manifold of low energy states [see Eq.
(\ref{2.44})]. Although this manifold obeys $SO(5)$ dynamical
symmetry, transitions to the highest singlet excitonic state does
not enter explicitly into the set of dynamical variables involved
in Kondo cotunneling. However the virtual transitions to this
state may also influence the properties of S/T pair.

This situation is especially important for tunnelling through DQD
with ${\cal N}=2$ (see \cite{KA01}). In this case the wave
functions of electrons in $S$ and $T$ states are different because
the intradot tunnelling $V$ intermixes two singlet states [$E_S$
and $E_R$ in Eq. (\ref{2.3}), $E_S$ and $E_L$ in Eq. (\ref{2.4})],
while the triplet $E_T$ is not affected by this process. As a
result, a difference in tunnelling rates arises like in the case
of planar QD, but in this case the sign of the difference is
strictly determined, namely $W^{T\sigma}>W^{S\bar\sigma}$.

In a generic situation,
one starts  from the original Anderson Hamiltonian
  (\ref{3.0a}), (\ref{3.2}),
(\ref{tun1}) and then arrives at the renormalized Kondo
Hamiltonian (\ref{cotron2}) with effective vertices ${\cal
J}_1,{\cal J}_2 $ by two steps \cite{Hald}. First, one has to take
into account renormalization of the bare energies $E_\Lambda$ in
${\cal H}_{dot}$ due to reduction of the energy scale $D_0\to D$.
This reduction is governed by the scaling equations
\begin{equation}\label{3.8}
dE_\Lambda/dL=-\Gamma_\Lambda/\pi D.
\end{equation}
The scaling trajectory for $E_\Lambda(D)$ is determined by a
scaling invariant \cite{Hewson,Hald}
\begin{equation}
E_\Lambda^*=E_\Lambda(D)-\pi^{-1}\Gamma_\Lambda\ln (\pi
D/\Gamma_\Lambda)
\end{equation}
Due to the above mentioned difference in tunnel matrix elements,
$\Gamma^T> \Gamma^S$ the flow trajectories $E_S(D)$ and $E_T(D)$
may intersect. The necessary conditions for such level inversion
in the situations illustrated by {Fig.  \ref{f.6}b,c} are
discussed in \cite{KA01}.

The physical reason for the level crossing is relatively
transparent: tunnelling from DQD to metallic leads induces
indirect electron exchange between two wells, and this "kinetic
exchange" favors parallel aligning of spins in each well. If this
"ferromagnetic" coupling
 overcomes antiferromagnetic indirect
coupling $2\beta_i V$ in (\ref{2.3}), (\ref{2.4}), the S/T
transition occurs due to accidental degeneracy, which stems from
the dynamical $SO(4)$ symmetry of ${\cal H}_{dot}$.

Apparently, S/T transition is a widespread phenomenon in the
 physics of quantum dots with even occupation. It was observed in
 vertical dots \cite{Sasa}, in lateral dots \cite{Wiel,Kyri,Kogan}, in
 quantum rings \cite{Ens04} and discussed in many theoretical
 studies \cite{taglia,Hofst02a,Hofst02b,Pust01b,Pust01a,Pust03}. One
 should note that the enhancement of Kondo temperature in the
 crossover region mentioned above is accompanied by more refined
 manifestations of the pertinent accidental degeneracies.
\begin{figure}[ht]
\includegraphics[width=8cm,angle=0]{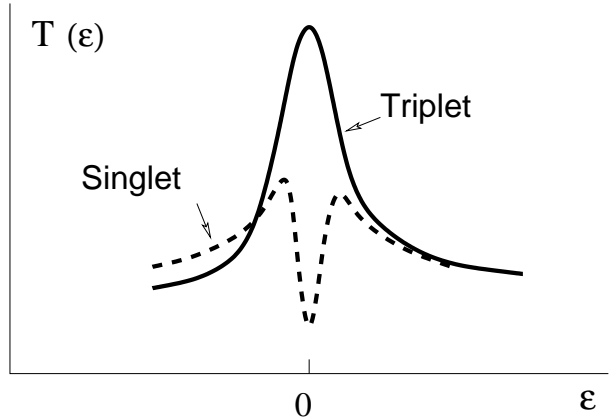}
\caption{Tunnelling transparency of QD with even occupation on
triplet (solid line) and singlet (dashed line) side of T/S
crossover. }\label{f.8}
\end{figure}

Indeed, on the triplet side of the Kondo cotunneling regime, one
deals with {\it underscreened} $S=1$ Kondo model \cite{nobl80},
where only half of the impurity spin is dynamically screened by
conduction electrons, provided the transformation (\ref{rot})
eliminates one of the two tunnel channels (otherwise the screening
is complete). The role of low-lying singlet excitation is
manifested by a modification of the Kondo temperature
$T_K(\Delta_s)$ [see Eq. (\ref{3.9b}) and discussion around]. On
the singlet side of this transition where $\Delta_s>0$, formation
of singlet ground state is a two-stage process \cite{Hofst02a}.
Until the characteristic energy $\varepsilon$ or temperature $T$
exceeds $\Delta_s>0$ the spin $S=1$ is still underscreened, Kondo
resonance in cotunneling enhances  the transmission coefficient
${\cal T}(\varepsilon)$, so the latter grows with decreasing
$\varepsilon$. This growth ceases at $\varepsilon \sim \Delta_s$,
and ${\cal T}(\varepsilon)$ falls off rapidly, because at zero
energy/temperature the Kondo regime is completely quenched. Thus,
the ZBA of the conductance should have a form of a narrow Kondo
peak like at the triplet side of the S/T transition, however, a
pronounced dip should emerge at low energies $\varepsilon<
\Delta_s$, so that conductance disappears at zero energy (see
Fig.\ref{f.8}). Such behavior was reported in \cite{Hofst02a} for
a DQD in the parallel geometry (Fig. \ref{f.6}b) employing methods
of numerical renormalization group (NRG) \cite{nrg}. Later on, a
similar behavior was substantiated for a planar QD with
anomalously small interlevel spacing $\delta\varepsilon$
\cite{Pust03}. In this case, the  intradot exchange $J_d$ nearly
compensates $\delta\varepsilon$ (see upper line in Eq.
\ref{gapp}), the gap $\Delta_s $ is close to zero and the S/T
transition may be induced by an external magnetic field applied
perpendicular to the plane of the dot. One may neglect the Zeeman
shift in comparison with orbital effects due to the small value of
the g-factor in GaAs . Consequently, the ZBA peak in the
conductance as a function of $\Delta_s$ should have a pronounced
maximum at $\Delta_s=0$ with asymmetric slopes at the singlet and
triplet sides of this crossover (Fig. \ref{f.9}). Apparently, all
above effects were observed in the experiment \cite{Wiel}.
\begin{figure}[ht]
\includegraphics[width=8cm,angle=0]{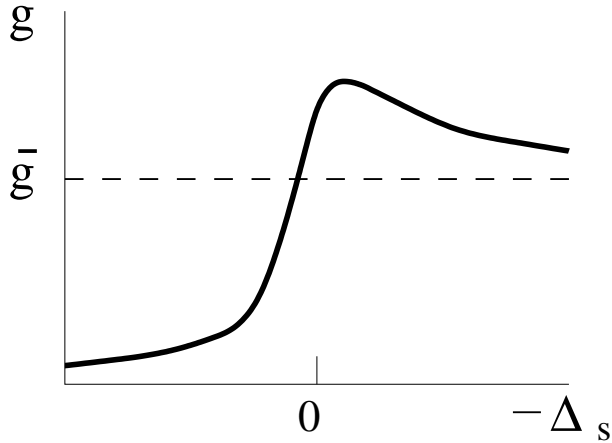}
\caption{Conductance of QD with even occupation as a function of
exchange gap $\Delta_s$ at fixed temperature $T$. $\bar g$ is a
limiting value of conductance for completely quenched singlet,
$|\Delta_s|\to \infty$.}\label{f.9}
\end{figure}

A two-stage Kondo effect may occur also on the triplet side of the
S/T transition in planar QD due to the difference between the
tunnelling amplitudes for the levels $\varepsilon_1$ and
$\varepsilon_2$ \cite{Pust01b}. In this case, one has two
tunneling channels for each lead characterized by the coupling
constants $W_{1l}, W_{1r}, W_{2l}, W_{2r}$. The $2 \times 2$
tunnelling matrix can be diagonalized by a transformation similar
to (\ref{rot}). Then one is left with two channels instead of
four. These two channels are sufficient for a complete screening
of the spin S=1. To describe the screening of a spin rotator under
these conditions, one may use the mathematical option of
representing the group $SO(4)$ as a direct product $SU(2)\times
SU(2)$. This can be carried out by a simple transformation
\cite{Mama,Eng,KA01,Pust01b}:
\begin{equation}\label{3.10}
{\bf S}= {\bf s}_1 +{\bf s}_2, ~~~{\bf R}= {\bf s}_1 -{\bf s}_2.
\end{equation}
Here ${\bf s}_{1,2}$ are two fictitious spins 1/2 operators. Then the
original problem can be mapped on the two-site Kondo problem,
where each spin is coupled to the reservoir with its own exchange
constant, $J_1\neq J_2$. Unlike the case encountered in the conventional
two-impurity Kondo effect \cite{Varma}, these fictitious spins are
not independent, but the RG equations still may be derived, and
the two-stage Kondo screening means that each channel is
characterized by its own Kondo temperature $T_K(J_i)$. As a
result, the unitarity limit is reached first in the channel with
larger $J_i$, so that half of the spin 1 is screened, and then the
remaining part of this spin is screened at lower temperatures. As
a result, the conductance versus temperature curve
acquires a stepwise form.

All these characteristics can be regarded as direct manifestations
of dynamical $SO(4)$ symmetry inherent in the S/T multiplet. The
picture becomes even richer when a parallel magnetic field is
applied. a Zeeman splitting has a special role in the context of
dynamical symmetry within the physics of Kondo cotunnelling
through QD. These features are discussed in the next subsection.
\begin{figure}[ht]
\includegraphics[width=13cm,angle=0]{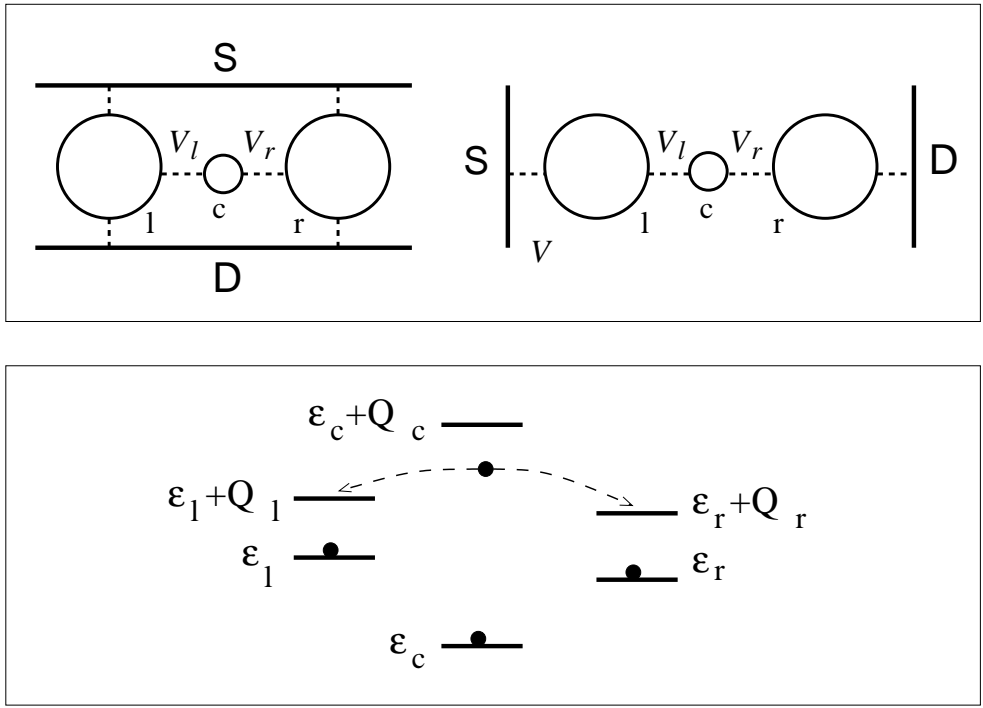}
\caption{Top panel: TQD in parallel and series geometry. Lower
panel: energy levels of isolated TQD occupied by four electrons.
The last electron is poised between the left and right
dot.}\label{f.10}
\end{figure}

More refined cases of accidental degeneracy within spin multiplets
arise in complex QD containing more than two valleys. One of such
examples, a triple quantum dot (TQD) was considered recently in
Refs. \cite{KKA}. One can arrange a TQD both in parallel and in
series geometry (see Fig. \ref{f.10}a,b). It possesses a richer
low-energy spectrum than DQD and planar QD structures considered
above. In particular, if the central dot has a smaller size (and
hence larger value of Coulomb blockade parameter $Q_c$), than the
side dots, and if the latter two are identical, then, beside other
features, the isolated TQD has an additional mirror symmetry.


Let us discuss the representative case of TQD with even occupation
${\cal N}_0=4$ ( see Fig. \ref{f.10}c for the energy level
scheme). Then three electrons occupy the levels
$\varepsilon_l,\varepsilon_c,\varepsilon_r$ and the last one is
shared between the states $\varepsilon_i+Q_i~(i=l.r)$ avoiding the
central well due to its high charging energy. When the inter-well
tunnelling amplitudes $V_{l,r}$ are taken into account, five
low-energy states determine the dynamical symmetry: these are two
singlets, two triplets and one charge transfer exciton. The mirror
symmetry of such TQD may be broken by the gate voltages applied to
the $l$ and $r$ wells. Then, the above mentioned Haldane-Anderson
two-stage RG procedure applied to the system TQD + leads results
in flow diagrams for these levels. Depending on the initial
conditions, one encounters various cases of accidental degeneracy,
which are connected with numerous $SO(n)$ groups. In the
completely symmetric case, the relevant group is $P\times
SO(4)\times SO(4)$ (two singlets and two triplets entering the
effective SW Hamiltonian). Here $P$ stands for the $l-r$
permutation operator describing the mirror symmetry of the TQD.
Another interesting cases which arise when the mirror symmetry is
broken are described by the familiar $SO(5)$ group (two singlets
and one triplet) and by a more exotic $SO(7)$ group (two triplets
and one singlet). The effective SW Hamiltonian for Kondo
cotunneling is an obvious generalization of (\ref{cotron}):
\begin{equation}
\label{cotron3} {\cal
H}_{cot}=\sum_{\alpha\beta}\left(J^{(1)}_{\alpha\beta} {\bf
S}_{\alpha\beta}\cdot{\bf s}_{_{\alpha\beta}}+
J^{(2)}_{\alpha\beta} {\bf R}_{\alpha\beta}\cdot{\bf
s}_{_{\alpha\beta}}\right).
\end{equation}
($\alpha,\beta=l,r)$. The number of vertices in this Hamiltonian
is predetermined by the number of vectors in the Fock space for
the $SO(n)$ group.
\begin{figure}[ht]
\includegraphics[width=12cm,angle=0]{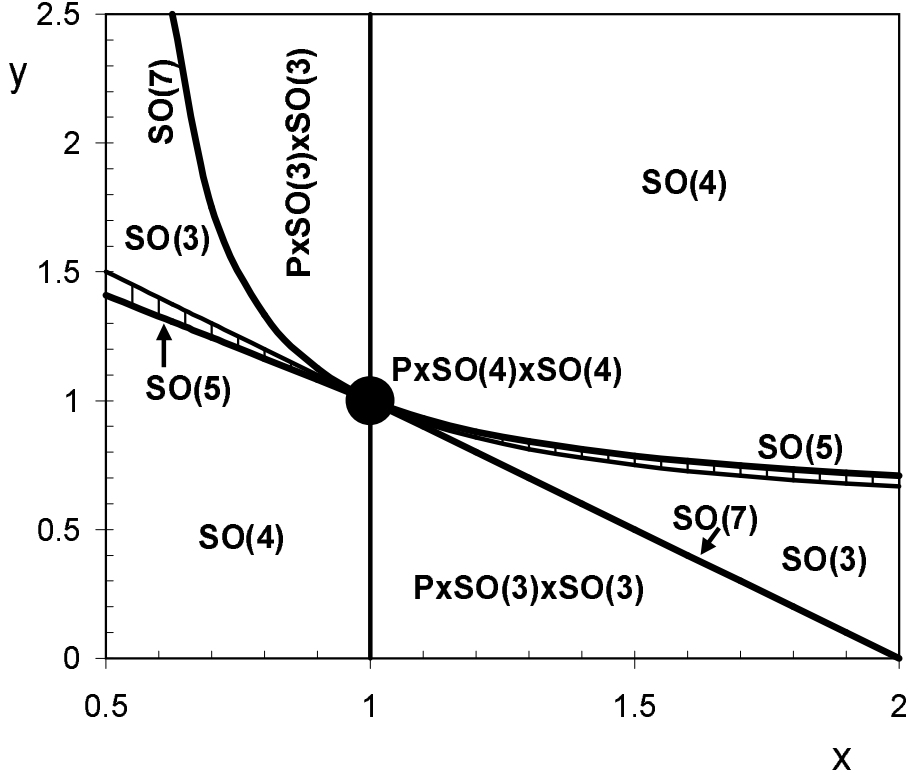}
\caption{Dynamic symmetries in TQD: Phase diagram in coordinates
$x=\Gamma_l/\Gamma_r$,
$y$$=$$($$\varepsilon_l$$-$$\varepsilon_c$$)$$/$$($$\varepsilon_r$$-$$\varepsilon_c$$)$.}\label{f.11}
\end{figure}

A general phase diagram of the TQD system is presented in Fig.
\ref{f.11}. One may find in Ref. \cite{KKA} the full list of group
generators and various scaling equations with solutions for $T_K$
specific for each dynamical group. As a result, in addition to the
S/T transition, which in this phase diagram is a transition
between the $SO(3)$ and the (shaded) singlet domains, one gets a
rich variety of other transitions and continuous crossovers, which
may be tested experimentally by changing gate voltages $v_{gi}$
and tunnel rates $\Gamma_i$ and observing the variation of the ZBA
peak as a function of these parameters and temperature.

\subsection{Kondo cotunnelling in an external magnetic field}

In the previous section we discussed the role of magnetic field as a
source of accidental level degeneracy due to diamagnetic shift
(\ref{2.66}) of single electron levels in QD. Here we discuss its
role as a source of Zeeman splitting. Zeeman effect directly
influences the symmetry properties of spin states because it
violates rotational invariance. Its contribution to conventional
Kondo effect is well known \cite{WT,AM,Melnik}. Due to lifted
Kramers degeneracy of the impurity  spin 1/2 state, the external
magnetic field splits the Abrikosov-Suhl resonance into two peaks and
eventually suppresses the Kondo effect, when the Zeeman splitting
$E_Z$ essentially exceeds $k_BT_K$. Dynamical symmetry inherent in
QD may radically change this relatively simple picture. The most striking
effect of magnetic field on quantum tunnelling through QD is a
{\it magnetic field induced Kondo effect}, which is absolutely
counterintuitive from the point of view of conventional Kondo
physics. This effect predicted in \cite{PAK} and immediately
confirmed experimentally \cite{ext} on single-wall nanotubes was
in fact the first explicit manifestation of dynamical symmetry in
Kondo tunnelling.
\begin{figure}[ht]
\includegraphics[width=6cm,angle=0]{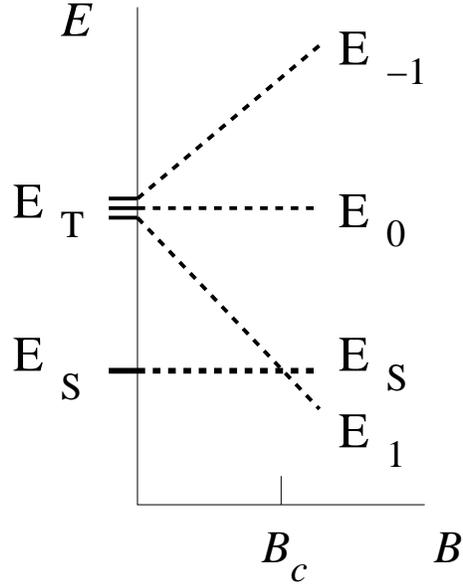}
\caption{DQD: accidental degeneracy in Zeeman field.}\label{f.12}
\end{figure}

The physical mechanism of magnetic field induced Kondo tunnelling
in QD with even occupation is rather understood. In the absence of
strong exchange forces, the ground state of such QD is spin
singlet (see above), and the lowest excited state is a triplet
(see Figs. \ref{f.3}c and \ref{f.6}). These two states are
separated by the spin excitation the gap energy $\Delta_s$
(\ref{gapp}). The Zeeman splitting energy $E_Z$ may compensate
this gap for the triplet state with spin up projection and thus
result in an accidental degeneracy of a spin multiplet (Fig.
\ref{f.12}). Mathematically, this means that the symmetry of a
non-compact group $SO(4)=SU(2)\times SU(2)$ is reduced to an
$SU(2)$ symmetry of the degenerate $\{S,T1 \}$ "doublet". This
reduction can be described in terms of generators of a dynamical
group \cite{KA01} by means of the vector operator ${\bf P}_1$ with
components,
\begin{equation}
P^+_1=\sqrt{2}X^{1S},~~~P_1^z=X^{11}-X^{00},
\end{equation}
which generates the $SU(2)$ subgroup describing the doublet
$E_S,E_{T1}$. The complementary vector operator ${\bf P}_2$ defined as
\begin{equation}
P^+_2=\sqrt{2}X^{0\bar 1},~~~P_2^z=X^{00}-X^{\bar 1 \bar 1},
\end{equation}
generates a second subgroup $SU(2)$. Then the effective Hamiltonian
describing Kondo cotunnelling induced by magnetic field in this
QD has the form,
\begin{equation}\label{cotronz}
H_{cotun}^Z= E_GP_0 +J^{(2)}{\bf P}_1\cdot {\bf s}~.
\end{equation}
This Hamiltonian should be compared
with that of equation (\ref{cotron}). Here
$P_0=X^{11}+X^{00}$ and $E_G=E_S=E_{T1}$ is the degenerate
ground-state energy level of a QD in a "resonant" Zeeman field
$B_c$. The Hamiltonian (\ref{cotronz}) describes Kondo-like
tunnelling, which arises at $E_z=\Delta_s$ and disappears once the
resonance conditions are not met. Precisely this effect was observed in
measurements of conductance through a segment of nanotube confined
by two metallic electrodes \cite{ext}. The Kondo-type ZBA in
the Coulomb windows with even N was absent at zero $B$, then suddenly
appeared at $B=B_c=1.38$ T and
afterward disappeared again at $B>B_c$.

A similar situation is possible in TQD \cite{KKA}. However, in
this more complicated structure, additional accidental approximate
degeneracy between two singlet levels, e.g., $E_{S}^{l}$ and
$E_{S}^{r}$ and one triplet state, e.g.,  $E_{T1}^{r}$ is
feasible. In this case the symmetry of TQD in magnetic field $B_c$
is $SU(3)$ [see Eqs. (\ref{1.4a}), (\ref{1.4})], Kondo tunnelling
violates this symmetry, and the corresponding cotunneling
Hamiltonian is {\it generically anisotropic}. Similar situation
arises when the Zeeman splitting induces crossing of two triplet
states $E_{T}^{l}$ and $E_{T}^{r}$ with one singlet state, say
$E_{S}^{r}$.

Another example of unusual sensitivity of conductance to Zeeman
effect in lateral QD with even occupation was discussed in
\cite{Pust03}. As was shown in Section II.A [see also discussion
around Eq. (\ref{3.10})], the conductance in this case is
predetermined by existence of two tunnelling channels, since two
single electron levels $\varepsilon_{1,2}$ are involved in
formation of S/T multiplet. Tunnelling through each of these
levels is characterized by the scattering phase shift
$\delta_{n\sigma}$ \cite{GR,Langr}, and the conductance can be
expressed via these phase shifts at the Fermi energy $(n=1,2)$.
For example, in a specific case $W_{l1}=-W_{r1},~W_{l2}=W_{r2}$
this expression is especially simple:
\begin{equation}
g(B) = \frac{1}{2}\sum_\sigma\sin^2
(\delta_{2\sigma}-\delta_{1\sigma}).
\end{equation}
Zeeman splitting results in repopulation of the levels
$\varepsilon_{n\sigma}$. In a singlet state $\delta_{1\sigma}=
\pi$, $\delta_{2\sigma}= 0$ for both spin projections $\sigma$. At
finite $B$ these shifts evolves under Friedel-Langer sum rule
constraint
$$
\sum_{n\sigma} \delta_{n\sigma}(B) =2\pi ,
$$
in compliance with the occupation number ${\cal N}=2$. As a result
of this repopulation the conductance as a function of the magnetic
field $B$ acquires a highly non-monotonous dependence with a
maximum occurring within the S/T crossover region.

To conclude this subsection, we emphasize that the main role of
magnetic field is that it leads to violation of rotational
symmetry in spin space. This violation results in the appearance
of magnetic anisotropy of Kondo cotunnelling observables and
non-trivial dynamical symmetries from $SU(n)$ family.

\section{Dynamical symmetries at finite frequencies}

Various manifestations of broken dynamical symmetry of quantum
dots in the Kondo regime discussed in Section III were related to
the low-energy sector of the spin excitation spectrum scaled by
the Kondo temperature $T_K$. In principle, one may imagine a
situation where the accidental degeneracy of energy levels arises
in some limited interval separated by finite energy from the
ground state of a nanoobject. Tunnelling into and out of electron
reservoirs at finite energies should then be described by an
effective Hamiltonians operating in the pertinent chosen interval
$\Delta\varepsilon$. An appropriate method for studying dynamical
properties of such systems is,  for example, functional RG
approach where a given Hamiltonian is diagonalized by continuous
unitary transformation \cite{fRG}.

However, if one intends to remain within a definite manifold of
discrete eigenstates of the QD, a frequency dependent perturbation
should keep the system in quasiequilibrium, where the response is
still determined by the thermodynamic partition function rather
than by essentially non-equilibrium distribution functions of the
electrons in the leads. This restriction includes the
experimentally important probe, that is, an interaction with a
monochromatic electromagnetic field.

Now the question is, what kind of dynamical symmetry might prevail
in photo-excited states, i.e., at finite excitation energy $h\nu$
of an absorbed light quantum. The main mechanism of light
absorption in QDs is formation of excitons. Internal hidden
symmetry of exciton spectra was briefly discussed in Subsection
II.C in connection with multiexciton absorption in semiconductor
nanoclusters. Here we focus our attention on dynamical effects
involving Kondo processes and address the problem of {\it whether
it is possible to observe Kondo effect at energies $h\nu\gg T_k$}.

The  Kondo effect in the presence od an alternating
electromagnetic field was intensively studied
during the last decade \cite{Hett}. In these studies, the
interest was focused on the interaction between spin degrees of
freedom of the QD on the one hand, and electron-hole
photo-excitations in the leads on the other hand. The main effect
of coherent light absorption in the leads is reflected through the
occurrence of higher Kondo harmonics in the tunnel current as a
response to the corresponding harmonics in the metallic band
continuum (see recent review \cite{plat} for general view on the
photon-assisted transport through QDs). Yet, from the point of
view emphasized here, this interesting effect is not related
directly to the dynamical symmetries of the QD. Indeed, the object
of our interest has to do with the manifold of eigenstates of the
QD, so we will discuss the Kondo aspects of excitonic states
created by resonance light absorption {\it in the quantum dot
itself}.

The problem of {\it dynamically induced Kondo effect} in QD with
even occupation $\cal N$ was formulated in Ref. \cite{KA00}. The
theory was addressed first of all to systems composed of
semiconductor nanoclusters (see Subsection II.C). The ground state
of such nanocluster is fully occupied valence states divided by
the energy gap $\Delta_g$ from the empty conduction states, so
that the occupation $N$ is always even. External visible light
with $h\nu\sim \Delta_g $ may create electron-hole pairs
(excitons)  but the occupation of a QD remains, of course, even.
The Hamiltonian responsible for creation of electron-hole pairs
has the form
\begin{equation}\label{Hex}
{\cal H}^{\prime}= \sum_{cv}\sum_{\sigma}D_{cv}d_{c\sigma
}^{\dagger }d_{v\sigma }\exp(-i\omega t)+ h.c. \label{2.0}
\end{equation}
Here $D_{cv}$ is the matrix element of the dipole operator
$\hat{D}$, while the indices $c,v$
stand for electrons in conduction and
valence levels, respectively . This term should be added to
the conventional Anderson Hamiltonian ${\cal H}_{dot}+{\cal H}_{tun}$.
To demonstrate the mechanism of dynamically induced Kondo
tunnelling, it is sufficient to retain in ${\cal H}_{dot}$ only the
ground state $|0\rangle$ and the lowest singlet exciton
\begin{equation}\label{Exx}
|E\rangle=\frac{1}{\sqrt{2}}\sum_{\sigma }d_{c\sigma }^{\dagger
}d_{v\sigma }|0\rangle,
\end{equation}
(triplet excitons remain dark in
the context of optical transitions). One
may then substitute the configuration changing operator $X^{E0}$ for
$d_{c\sigma }^{\dagger }d_{v\sigma }$ in the Hamiltonian
(\ref{Hex}). The tunnel Hamiltonian intermixes the states
$|0\rangle$ and $|E\rangle$ with the continuum states
\begin{equation}
|kc\rangle =\frac{1}{\sqrt{2}}\sum_{\sigma }d_{c\sigma }^{\dagger}
c_{k\sigma }|0\rangle ,\;\;
|kv\rangle =\frac{1}{\sqrt{2}}%
\sum_{\sigma}c_{k\sigma }^{\dagger } d_{v\sigma }|0\rangle .
\label{contex}
\end{equation}
Using these definitions, ${\cal H}_{tun}$ can be written in terms
of new configuration changing operators,
\begin{equation}
H_t=V_c\sum_k\left(|kc\rangle\langle 0|
+\frac{1}{\sqrt{2}}|kv\rangle\langle E|\right) + V_v\sum_k
\left(|kv\rangle\langle 0| -\frac{1}{\sqrt{2}}|kc\rangle\langle
E|\right) + H.c. \label{tunex}
\end{equation}

The building blocks for the secular equation, which determine the
response function $R(\epsilon)$
\begin{equation}
R(\epsilon)=\langle 0\left|\hat D\frac{1}{\epsilon-{\cal H}} \hat
D\right|0\rangle,
\end{equation}
are shown in {Fig. \ref{f.13}}. One immediately recognizes the two
intermediate states as the first terms of a Kondo series for a
state with one excess electron and one excess hole  in the QD. To
lowest order in $V_{c,v}$, this secular equation reads,
\begin{equation}
\det \left|
\begin{tabular}{cc}
$\epsilon -\Sigma_{00}(\epsilon)$ & $\quad -\Sigma _{E0}(\epsilon)$ \\
$-\Sigma_{0E}(\epsilon)\quad $ & $\epsilon -\Delta -\Sigma
_{EE}(\epsilon)$%
\end{tabular}
\right| =0. \label{2.10}
\end{equation}
The real parts of the self energies contain the familiar Kondo
logarithmic divergencies
$$
{\rm Re}\Sigma_{jl}(\epsilon)\sim \frac{\Gamma_{jl}}{2\pi }\ln
\frac{(\epsilon -\Delta_{c,v})^2+(\pi T)^2}{D^2},
$$
with   $\Gamma_{jl}=\pi \rho _0 V_j^*V_l.$ These functions have
sharp maxima at energies
$\Delta_{c,v}=\Delta_g-\varepsilon_{c,v},$ which can be considered
as "precursors" of Kondo peaks in a weak coupling regime $T\gg
T_K$.
\begin{figure}[ht]
\includegraphics[width=11cm,angle=0]{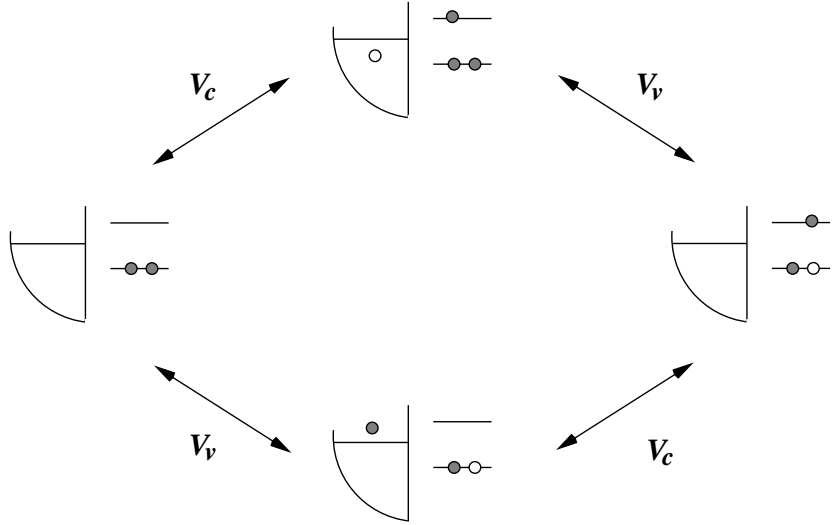}
\caption{Building blocks of secular equation (62). }\label{f.13}
\end{figure}

The secular equation (\ref{2.10}) describes a dynamical mixing
between the states $|0\rangle$ and $|E\rangle$ with maxima at
finite frequencies $h\nu\sim\Delta_{c,v}\gg T_K$ within the
multiplet of eigenstates of the QD. This scenario has no analogs
in our previous examples of broken dynamical symmetry. Here, the
Kondo processes arise only in the intermediate states of light
absorption,  when either 1 or 3 electrons occupy the relevant
levels of the QD ({Fig. \ref{f.13}), whereas there is no Kondo
screening in the initial and final states of exciton absorption
due to the even occupation ${\cal N}_i$ and ${\cal N}_f$. The
above maxima of the self energies should be observed as satellite
peaks in optical lineshape $\sim {\rm Im}~R(h\nu)$ at energies
$\Delta_{c,v}$, which accompany the main peak of exciton
absorption at $h\nu \sim \Delta_g$. In spectroscopic nomenclature
these midgap excitonic states may be regarded as "Kondo shake-up
peaks".

An important constraint is imposed on dynamical Kondo screening by
the finite lifetime of electrons and holes in the photoexcited QD.
Indeed, too short  photo-electron, photo-hole or photo-exciton
lifetimes, $\tau_l<\hbar (k_BT_K)^{-1}$, prevents even the
formation of a precursor of the Kondo resonance, not to mention
the impossibility of reaching the strong coupling Kondo regime. To
ensure conditions for the observation of shake-up Kondo
satellites, one should design the heterostructure lead/dot/lead in
such a way that at least one of the two carriers in the QD has a
negligible tunnelling width (we refer the reader to Ref.
\cite{KA00} for a detailed analysis of this problem).
The problem of finite lifetime for excited Kondo states will be
addressed in more details in the next section.

In principle, dynamically induced Kondo effect may arise not only
in excitonic spectra, but also in photoemission spectra, where the
electron leaves the QD for the metallic lead under light
irradiation \cite{Fuji01}. In this case only two blocks from Fig.
\ref{f.13} are involved in dynamical mixing, namely the ground
state $|0\rangle$ and the state with excess hole $|kv\rangle$ in a
QD. This mixing  manifests itself as an edge singularity in the
absorption spectrum
$$
{\rm Im}~R(h\nu) \sim (h\nu - \Delta_{cv})^\alpha
$$
where the power $\alpha$ is predetermined by the phase shift of
electrons on the Fermi level of the lead. In case of inverse
process, when the excess electron is captured by the dot with
light quantum emission, the state $|kc\rangle$ is involved in
dynamical mixing. Of course, all these shake-up processes go back
to the celebrated edge-singularity in X-ray absorption spectra
discovered theoretically in mid sixties \cite{Mahan}.

The dynamical mixing processes discussed above arise in linear
response to external electromagnetic field. It is clear that
similar processes may arise also in non-linear optical response,
but the variety of possibilities is richer in this case. One such
possibility, which may be realized in pump-probe
spectroscopy was discussed in Ref. \cite{third}. According to
the proposal outlined therein,
a strong monochromatic pumping field with
frequency $\nu_1$ modifies the position of
the deep electron level in the
dot, $\varepsilon_d \to \varepsilon_d + h\nu_1$, and induces a
Kondo-resonance on the Fermi level described by the
self-consistent equation
$$
\epsilon=\Sigma(\epsilon)
$$
(cf. Eq. (\ref{2.10}). The quantity $\pi \rho_0({\bar d}{\cal
E}_1)^2$ can be interpreted as
 "tunnelling width'' $\Gamma$ in the self
energy $\Sigma(\epsilon)$ ($\bar d$ is the dipole moment and
${\cal E}_1$ is the amplitude of the probe electric field). On top
of it one should measure the probe light absorption. The analysis
of third order pump-probe polarization shows that when the
resonance condition $\nu_1\simeq \nu_2$ is valid, a dynamical
mixing between the pumped level $E_d=\varepsilon_d + h\nu_1$ and
the Kondo resonance at the absorption threshold
$h\nu_2=\varepsilon_F-\varepsilon_d$ arises. As a result, a narrow
shake-up satellite appears  below the absorption threshold.

One more peculiar possibility of Kondo shake-up processes for {\it
charged} excitons was discussed in Ref. \cite{govor}. Charged
exciton is a multiparticle complexes consisting of several valence
holes and conduction electrons trapped in a self-consistent
potential well. Such excitons containing one hole and $n+1$
electrons are observed in self-assembled InAs/GaAs quantum dots
discussed in Section II.C \cite{Warb00}. These complexes are
usually labelled as $X^{n-}$. The surface of wetting GaAs layer
may be filled with 2D electrons, and tunnelling between the trap
and the 2D electron continuum is possible. Some of these excitons
have non-zero spin, so the possibility of Kondo screening opens
up. According to Ref. \cite{govor}, the best candidate for Kondo
screening is the $X^{2-}$ exciton, which possesses nonzero spin
and sufficiently long lifetime necessary to form a screening Kondo
cloud. This exciton consists of a hole in an $s$ state and three
electrons (two of them are in $s$-states and one in a $p$-state).
Electron tunnelling between the trap and the 2D electrons in the
wetting layer admixes the state $|X^{3-}c_{ks}|0\rangle$ with an
additional electron captured into the trap to the state
$|X^{2-}|0\rangle$ in analogy with Eqs.
(\ref{contex}),(\ref{tunex}). The narrow  shake-up peak
 arises in the photoluminescence spectrum as a result of Kondo
 screening like in the cases discussed above.
 Hybridization between these states induced by photon emission
 was registered experimentally
 in photoluminescence spectrum measured in a strong magnetic
 field, where the 2D electronic states become discrete due to
 Landau quantization, and the excitonic levels are subject to
 diamagnetic shift \cite{Warb04}. Of course, there is no room for
 Kondo effect in case of fully discrete spectrum, but the dynamical
 hybridization seems to become an established fact.

\section{Dynamical symmetries in  non-equilibrium state}

Unlike the problem of magnetic scattering in doped metals, where
the Kondo effect in a strong coupling regime  is described in
terms of equilibrium thermodynamics both in the strong coupling
($T\ll T_K$) and in the weak coupling ($T > T_K$) limits, Kondo
tunnelling through QD may occur also under non-equilibrium
conditions \cite{noneq}. We discussed  in Section IV the special
case of Kondo effect under the influence of a monochromatic ac
field, which is applied either on the leads or to the QD. Here we
turn our attention to the case, where equilibrium conditions are
violated by the finite bias applied on the metallic leads.

There is an extensive literature on the Kondo tunnelling through QD
under non-equilibrium conditions. Referring the reader to recent
papers \cite{neq}, where this problem is discussed in
details, we concentrate here only on the manifestations of
dynamical symmetry violation in non-equilibrium Kondo effect. Such
a possibility arises when the S/T transitions {\it induced by
finite bias} are involved in Kondo cotunneling \cite{KEF}.

To be concrete let us consider a T-shaped double quantum dot (Fig.
\ref{f.5}c) as an example of an experimental setup, where an {\it
electric field induced Kondo effect} can be realized. The S/T
energy gap $\Delta_s$ in a T-shaped DQD (\ref{gapp}) is controlled
by the height and width of the barrier separating the two dots.
The tunnelling amplitude $W$ between the DQD and  the metallic
contacts is controlled by gate voltages generating the
corresponding barriers. If the total number of electrons ${\cal
N}=2n$ is even and $\Delta_s\gg T^{eq}_K$, where $T_K^{eq}$
characterizes the equilibrium Kondo temperature of an
underscreened $S=1$ Kondo effect, the ground state of the dot is a
singlet and there is no Kondo enhancements of the differential
conductance associated with co-tunnelling processes in equilibrium
(infinitesimally small voltage between source and drain). However,
a finite bias applied to the leads changes this situation
 dramatically . The electrons, accelerated
by the bias $v_b$, gain an additional energy $\sim ev_b$ which may
compensate the energy $\Delta_s$ necessary for spin flips, and
thus make Kondo co-tunnelling possible (see Fig. \ref{f.14} where
the processes involved in cotunneling are shown by dashed lines).
Recalling that the pertinent charge conserving system is not the
isolated dot, but, rather, the closed circuit composed of
source-dot-drain, we may draw the conclusion that the
non-equilibrium Kondo tunnelling, which develops in a "moving
frame" of a system with dynamical symmetry results in a finite
bias anomaly (FBA) in contrast with the ubiquitous ZBA
characteristic for the conventional (equilibrium) Kondo effect.
\begin{figure}[ht]
\includegraphics[width=8cm,angle=0]{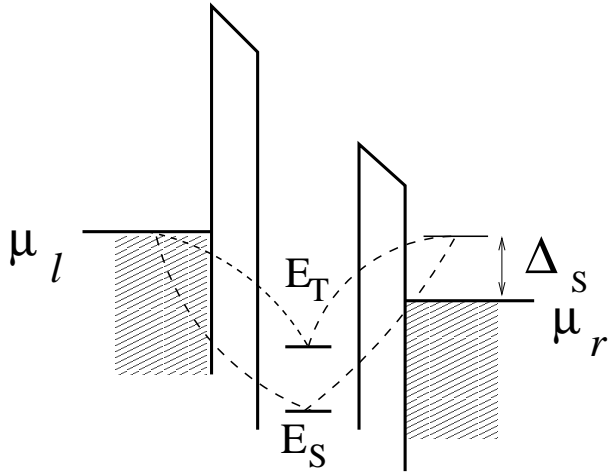}
\caption{DQD at finite bias.}\label{f.14}
\end{figure}

To characterize non-equilibrium Kondo effect in DQD, we adopt the
$SO(4)$ Kondo Hamiltonian (\ref{cotron3}). This Hamiltonian can be
derived for nonzero bias with the help of out of equilibrium SW
transformation \cite{kng}. As a result, at small bias (linear
response regime) the exchange integrals
$J_{\alpha\alpha'}^{\Lambda\Lambda'}$ become time-dependent, but
the scaling equations can be derived from the condition of the
invariance of the linear conductance. However, the Hamiltonian
(\ref{cotron3}) may be derived in a different fashion assuming
that both leads remain in quasi equilibrium and possess their own
chemical potentials $\mu_{L,R}$ under the constraint
$\mu_L=\mu_R+eV$. The relaxation processes in the leads are
negligible provided the leads are sufficiently hot and
phonon-assistant thermalization processes are fast enough compared
with other characteristic time scales involved in this analysis.
This fact allows us to adopt an RG procedure for frequency -
dependent vertices (\cite{neq}). However, the dynamical symmetry
of the problem implies a possible further simplification of the RG
scheme. The key distinction between the formalism suggested in
Ref. (\cite{KEF}) for DQD and the Keldysh type description of an
$s=1/2$ dot in an external magnetic field under non-equilibrium
condition \cite{neq,hool} is this: In the first case, the presence
of an additional adjustable parameter, namely $|\Delta_s - eV|$
justifies the {\it standard} RG scheme if the resonance condition
$|\Delta_s - eV|\ll T\ll T^{eq}_K$ is fulfilled. Therefore, under
non-equilibrium conditions illustrated by Fig. \ref{f.14}, the
dynamical symmetry backs the quasi-equilibrium RG approach
generally inapplicable to conventional  Kondo effect out of
equilibrium.

As a result, the
equations for tunnelling vertices
$J_{\alpha\alpha'}^{\Lambda\Lambda'}$ at finite temperatures are
written as follows:
\begin{equation}
\frac{d {\cal J}^T_{ll}}{d L}=-\rho_0 ({\cal J}_{ll}^T)^2,\;\;\; \frac{d
{\cal J}^{ST}_{ll}}{d L}=-\rho_0 {\cal J}_{ll}^{ST} {\cal J}_{ll}^T, \label{ll}
\end{equation}
$$
\frac{d {\cal J}^T_{lr}}{d L}=-\rho_0 {\cal J}_{ll}^T {\cal J}_{lr}^T,\;\;\;
\frac{d {\cal J}^{ST}_{lr}}{d L}=-\rho_0 {\cal J}_{ll}^{ST} {\cal J}_{lr}^T,
$$
\begin{equation}
\frac{d {\cal J}^S_{lr}}{d L}=
\frac{1}{2}\rho_0\left({\cal J}_{ll,+}^{ST}{\cal J}_{lr,-}^{TS}
+\frac{1}{2}{\cal J}_{ll,z}^{ST}{\cal J}_{lr,z}^{TS}\right) \label{rg}
\end{equation}
(cf. Eq. \ref{3.7}).
The solution of the system of equations
(\ref{rg}) supplemented by the boundary
condition ${\cal J}_{\alpha\alpha'}^
{\Lambda\Lambda'}(D_0)=J^{\Lambda\Lambda'}$ are given by
$$
{\cal J}^T_{\alpha,\alpha'}=\frac{J^T}{1-\rho_0
J^T\ln(D/T)},\;\;\;
{\cal J}^{ST}_{\alpha,\alpha'}=\frac{J^{ST}}{1-\rho_0
J^T\ln(D/T)},
$$
\begin{equation}
{\cal J}^S_{lr}=J^S-\frac{3}{4}\rho_0
(J^{ST})^2\frac{\ln(D/T)}{1-\rho_0 J^T\ln(D/T)}. \label{srg}
\end{equation}
and determine a new non-equilibrium Kondo temperature $T^{neq}_K$
$\sim $ $D\exp(-1/\rho_0 J^T)$ $\sim $ $(T^{eq}_K)^2/D$. This
temperature should be compared with the relaxation rate
$~\tau_{ST}^{-1}$ determined by singlet-triplet transitions in the
dot. The condition $\hbar/\tau_{ST} \ll T_K^{neq}$ defines a
stability domain for the finite bias anomaly against the decoherence
processes associated with effects of re-population of the dot by
the finite bias. For realistic dots, the estimate
$\hbar/\tau_{ST}\sim (\Delta_s)^3/D^2$ holds (see \cite{KEF}),
 and the Kondo effect is robust in a broad
domain of external parameters. This means that FBA arises at
$eV\approx \Delta_s$ instead of conventional Kondo ZBA (see Fig.
\ref{f.15}).
\begin{figure}[ht]
\includegraphics[width=8cm,angle=0]{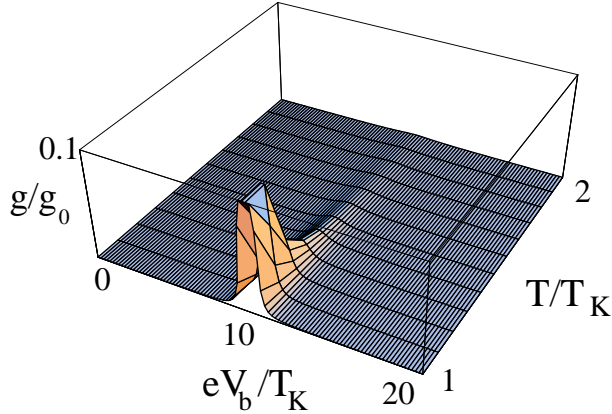}
\caption{Finite bias anomaly (FBA) in double quantum dot. Here
$g_0=e^2/\pi\hbar$ is the unitarity limit for quantum conductance,
which may be achieved at $T=0$ under condition of perfect Kondo
screening.}\label{f.15}
\end{figure}

The perturbative corrections corresponding to re-population
effects result in asymmetry of the FBA in the differential
conductance. The origin of this asymmetry (threshold character of
the relaxation associated with transitions between different
components of the triplet state) is however different from that
for the conventional non-equilibrium Kondo effect \cite{neq} and
is also attributed to the hidden dynamical symmetries associated
with the $SO(4)$ Kondo problem.

An interesting question which arises here is whether the
non-equilibrium Kondo effect falls into the class of
strong-coupling regime. It has been extensively studied during the
last few years (see discussion and references in \cite{neq,hool}).
The same question when addressed to systems characterized by
hidden dynamical symmetries allows a simple and straightforward
answer: the strong coupling limit is not achievable in this
situation. There is always an energy scale determined by an
external bias, decoherence effects associated with AC or effects
related to repopulation of the dot which prevent the system from
both one-stage and two-stage Kondo scenario \cite{Hofst02a,Pust03}
and suppress the Kondo effect in the ground state.

\section{Addendum. Dynamical algebras}

This chapter is a mathematical supplement to Chapter I. Here we
clarify the origin of spin dynamical symmetries and describe
several useful bosonization and fermionization procedures for the
generators of $SO(n)$ groups.

To demonstrate specific properties of dynamical algebras and
various representations of groups possessing hidden dynamical
symmetries we consider a group of 4-dimensional rotations. We
start with referring to the  well known description of a hidden
symmetry implicit in {\it spherical quantum mechanical rotator}.
The Hamiltonian of rotator is
\begin{equation}\label{a.1}
{\cal H}_r=\frac{1}{2I}{\bf L}^2
\end{equation}
where ${\bf L}$ is the operator of orbital moment and $I$ is the
moment of inertia.

Lie algebras $o(n)$ are defined  on a basis of $\frac{1}{2}n(n-1)$
operators of infinitesimal rotation
\begin{equation}
D_{\alpha \beta}=-D_{\beta\alpha}=x_\beta \frac{\partial}{\partial
x_\alpha}-x_\alpha \frac{\partial}{\partial x_\beta},
\;\;\;\;\;(\alpha<\beta=1,2,...n)\label{DD}
\end{equation}
which possess the following commutation relations
\begin{equation}
[D_{\alpha\beta} D_{\mu\nu}]=(\delta_{\alpha\mu}
D_{\beta\nu}-\delta_{\alpha\nu} D_{\beta\mu}-\delta_{\beta\mu}
D_{\alpha\nu}+\delta_{\beta\nu} D_{\alpha\mu}).\label{cc1}
\end{equation}
(see, e.g., \cite{Eng,Wyb}. The antisymmetric tensor
$D_{\alpha\beta}$ for $o(4)$ algebra acts in 4-dimensional space.
It can be parametrized in terms of two vectors $\bf {L}$ and $\bf
{M}$ as follows
\begin{eqnarray}
-i \left(
\begin{array}{cccc}
  0& L_3& -L_2 & M_1 \\
  &  0& L_1& M_2 \\
  & & 0& M_3 \\
  & & & 0
\end{array}
\right)
\end{eqnarray}
where the infinitesimal operators of $SO(4)$ group \cite{Wyb} in
4-dimensional space $(x,y,z,t)$ are given by
\begin{eqnarray}
L_1&=&i \left(y\frac{\partial}{\partial
z}-z\frac{\partial}{\partial y}\right),\;\; L_2=i\left(
z\frac{\partial}{\partial x}-x\frac{\partial}{\partial
z}\right),\;\; L_3=i \left(x\frac{\partial}{\partial
y}-y\frac{\partial}{\partial
x}\right),\;\;\nonumber\\
M_1&=& i\left(t\frac{\partial}{\partial
x}-x\frac{\partial}{\partial t}\right),\;\; M_2= i\left(
t\frac{\partial}{\partial y}-y\frac{\partial}{\partial
t}\right),\;\; M_3=i \left(t\frac{\partial}{\partial
z}-z\frac{\partial}{\partial t}\right).
\end{eqnarray}
From here we deduce the commutation relations:
\begin{equation}
[L_j,L_k]  =
ie_{jkl}L_l,\;\;\;\;\;\;[M_j,M_k]=ie_{jkl}L_l,\;\;\;\;\;\;\;\;\;[M_j,L_k]=ie_{jkl}M_l.
\label{com1}
\end{equation}
Six  generators of $SO(4)$ group are defined in a in 4-dimensional
spherical coordinate system characterized by 3 angles
$\theta,\phi, \alpha$,
\begin{eqnarray} x_1&=&x=R
\sin\alpha\sin\theta\cos\phi\;\;\;\;
x_2=y=R\sin\alpha\sin\theta\sin\phi \nonumber\\
x_3&=&z=R\sin\alpha\cos\theta\;\;\;\;\;\;\;\;\;\;\;\;
x_4=t=R\cos\alpha.
\end{eqnarray}

Returning back to spherical rotator, one may treat angle $\alpha$
as an angle between the rotation axis and fixed Cartesian z-axis.
Two rest angles characterize rotation in a 3D subspace with fixed
rotation axis. The operators of infinitesimal rotations are is
given by
$$
L_3=-i\frac{\partial}{\partial \phi},\;\;\;\;\;\; L^{\pm}=e^{\pm
i\phi}\left(\pm \frac{\partial}{\partial \theta}+i
\cot\theta\frac{\partial}{\partial \phi}\right),
$$
$$
M_3=i\cos\theta\frac{\partial}{\partial
\alpha}-i\sin\theta\cot\alpha\frac{\partial}{\partial \theta},
$$
$$
M^\pm=e^{\pm
i\phi}\left(\cot\alpha\left[\mp\frac{1}{\sin\theta}\frac{\partial}{\partial
\phi}+i\cos\theta\frac{\partial}{\partial
\theta}\right]+i\sin\theta\frac{\partial}{\partial \alpha}\right),
$$
where ladder operators $L^\pm =L_1\pm i L_2$ and  $M^\pm =M_1\pm i
M_2$. The angular moment $\bf{L}$ is parameterized by only two
angles, $\theta,\phi$ according to standard representation of
$SO(3)$ group \cite{Landau3}. Although the Hamiltonian (\ref{a.1})
contains only the invariant ${\bf L}^2$, describing 3D rotation,
the hidden 4th dimension is accessible due to additional generator
$\bf{M}$, which determines the transitions from 3D space to the
true 4D manifold \cite{assym}.
 As a result the operator
$\bf{L}^2$ is no more conserving quantity for $SO(4)$ group. The
Casimir operator
\begin{equation}\label{casim}
\bf{L}^2+\bf{M}^2=-\frac{\partial^2}{\partial
\alpha^2}-2\cot\alpha\frac{\partial}{\partial \alpha}-{\rm
cosec}^2\alpha\left(\frac{\partial^2}{\partial\theta^2}+
\cot\theta\frac{\partial}{\partial\theta}+{\rm
cosec}^2\theta\frac{\partial^2}{\partial \phi^2}\right)=3
\end{equation}
and the orthogonality condition
\begin{equation}\label{ortho}
\bf{L}\cdot\bf{M}=\bf{M}\cdot\bf{L}=0
\end{equation}
determine the hyperspherical harmonics as the eigen functions of
the angular part of 4-D Laplacian,
$$
Y_{nlm}(\alpha,\theta,\phi)=i^{n-1-l}2^{l+1}l!\left[\frac{n(n-l-1)!}{2\pi(n+l)!}\right]^{1/2}\sin^l\alpha
C_{n-l-1}^{l+1}(\cos\alpha)Y_l^m(\theta,\phi).$$ Here
$C_{n-l-1}^{l+1}$ is Gegenbauer polynomial \cite{abram} and
$Y_l^m(\theta,\phi)$ stands for standard  three dimensional
spherical harmonics.

It is known, that three generators of $SU(2)$ group together with
the Casimir operator $\bf{L}^2$ define a sphere $S_2$ where each
state is parameterized by two angles. The coherent states of
$SU(2)$ group may be constructed \cite{Per} by making a standard
stereographical projection of the sphere from its south pole to
the complex plane $z$. The space of generators of $SO(4)$ group is
6-dimensional, while 2 constraints determine 4-dimensional
surface, where each state is characterized by four angles. The
stereographical  projection of this surface on a 4-D complex
hyperplane  allows to construct coherent states for $SO(4)$ group.

The commutation relations (\ref{com1}) can be transformed into
another form by making the linear transformation to the basis
\begin{equation}\label{comb}
J_i=\frac{L_i+M_i}{2},\;\;\;\; K_i=\frac{L_i-M_i}{2}
\end{equation}
giving more simple commutation relations
\begin{equation}
[J_j,J_k]  =
ie_{jkl}J_l,\;\;\;\;\;\;[K_j,K_k]=ie_{jkl}K_l,\;\;\;\;\;\;\;\;\;[K_j,J_k]=0.
\label{com2}
\end{equation}
The operators $L_i, M_i$ as well as $J_i, K_i$  form the elements
of the Lie algebra $o(4)$. The operators $(J_1,J_2,J_3)$ and
$(K_1,K_2,K_3)$ are separately closed under commutations, each
describing a subalgebra of $o(4)$, namely $o(3)=u(2)$. The Lie
algebra $o(4)$ is the direct sum of two $o(3)$ algebras. This
splitting of the $o(4)$ algebra into two $o(3)$ subalgebras is
directly associated with the local isomorphism between the Lie
group $SO(4)$ with the direct product group $SU(2)\times SU(2)$.
The triads $(J_1,J_2,J_3)$ and $(K_1,K_2,K_3)$ each form proper
ideals \cite{Wyb} in $o(4)$, and the Lie algebra $o(4)$ is
semi-simple.

The symmetry group of {\it spin rotator} is defined in close
analogy with the above construction, but all rotations are
performed in a spin space. The triplet/singlet pair is formed in a
simplest case by two electrons represented by their spins $s=1/2$.
 Let us  denote them as $\vec{s}_1$ and $\vec{s}_2$. The components
of these vectors obey the commutation relations
\begin{equation}
[s_{1j},s_{1k}]  =
ie_{jkl}s_{1l},\;\;\;\;\;\;[s_{2j},s_{2k}]=ie_{jkl}s_{2l},\;\;\;\;\;\;\;\;\;[s_{1j},s_{2k}]=0
\label{com3}
\end{equation}
In similarity with (\ref{com2}) these vectors may be qualified as
generators of $o(4)$ algebra, which represents a  spin rotator.
Then, the linear combinations
\begin{equation}
S_i=s_{1i}+s_{2i},\;\;\;\; R_i=s_{1i}-s_{2i}
\end{equation}
are introduced in analogy with (\ref{comb}), which define 6
generators of SO(4) group possessing the commutation relations
(\ref{comm1}). These generators are represented in terms of the
Pauli-like matrices as follows
\begin{eqnarray}
S^+=\sqrt{2} \left(
\begin{array}{cccc}
  0& 1& 0 & 0\\
  0& 0 & 1& 0 \\
  0& 0& 0& 0 \\
  0& 0& 0& 0
\end{array}
\right),\;\;\;\; S^-=\sqrt{2} \left(
\begin{array}{cccc}
  0& 0& 0 & 0\\
  1& 0 & 0& 0 \\
  0& 1& 0& 0 \\
  0& 0& 0& 0
\end{array}
\right), \nonumber
\end{eqnarray}
\begin{eqnarray}
 S^z=\left(
\begin{array}{cccc}
  1& 0& 0 & 0\\
  0& 0 & 0& 0 \\
  0& 0& -1& 0 \\
  0& 0& 0& 0
\end{array}
\right),\;\;\;\; R^z=-\left(
\begin{array}{cccc}
  0& 0& 0 & 0\\
  0& 0 & 0& 1 \\
  0& 0& 0& 0 \\
  0& 1& 0& 0
\end{array}
\right),
\end{eqnarray}
\begin{eqnarray}
R^+=\sqrt{2} \left(
\begin{array}{cccc}
  0& 0& 0 & 1\\
  0& 0 & 0& 0 \\
  0& 0& 0& 0 \\
  0& 0& -1& 0
\end{array}
\right),\;\;\;\; R^-=\sqrt{2} \left(
\begin{array}{cccc}
  0& 0& 0 & 0\\
  0& 0 & 0& 0 \\
  0& 0& 0& -1 \\
  1& 0& 0& 0
\end{array}
\right). \label{matrix}
\end{eqnarray}
where the ladder operators $S^\pm = S^x \pm iS^y$, $R^\pm =R^x \pm
iR^y$. The constraints ({\ref{casim}}), (\ref{ortho}) now acquire
the form
$$
{\bf S\cdot R}=0,\;\;\;\;\;\; {\bf S}^2+ {\bf R}^2=3.
$$
 By construction $\bf{S}$
is the operator of the total spin of pair, which can take values
$S=0$ for singlet and $S=1$ for triplet states. The second
operator $\bf{R}$ is responsible for transition between singlet
and triplet states. Thus we come to the dynamical group $SO(4)$
for spin rotator introduced in Chapter I [see Eqs. (\ref{m.1}) and
(\ref{comm1})].

Similar procedure is used for the $SO(5)$ group. The corresponding
$o(5)$ algebra has 10 generators
$D_{\alpha\beta}=-D_{\beta\alpha}$ (\ref{DD}) satisfying
commutation relations (\ref{cc1}). These 10 generators may be
identified as 3 vectors and a scalar in a following fashion
\begin{eqnarray}
-i \left(
\begin{array}{ccccc}
  0& S^z& -S^y & R^x & P^x \\
  &  0& S^x& R^y & P^y \\
  & & 0& R^z & P^z \\
  & & & 0 & A\\
  & & & &0
\end{array}
\right)\label{s051}
\end{eqnarray}
where the operators $\bf{S},\bf{R},\bf{P}$ and the scalar operator
$A$ obey the following commutation relations
\begin{eqnarray}
\left[S_j,S_k\right]&=&
ie_{jkl}S_l,\;\;\;\;\;\;[R_j,R_k]=ie_{jkl}S_l,\;\;\;\;\;\;\;\;\;[P_j,P_k]=ie_{jkl}S_l,\nonumber\\
\left[R_j,S_k\right]&=&
ie_{jkl}R_l,\;\;\;\;\;\;[P_j,S_k]=ie_{jkl}P_l,\;\;\;\;\;\;\;\;\;[R_j,P_k]=i\delta_{jk}T,\nonumber\\
\left[P_j,A\right]&=&
iR_l,\;\;\;\;\;\;\;\;\;\;\;\;[A,R_j]=iP_j,\;\;\;\;\;\;\;\;\;\;\;\;\;\;\;\;\;
[A,S_j]=0. \label{com5}
\end{eqnarray}
[cf. Eqs. (\ref{comm1}), (\ref{comm2})]. The  operators $\bf{R}$
and $\bf{P}$ are orthogonal to $\bf{S}$, while the Casimir
operator is ${\cal K}={\bf S}^2+{\bf R}^2+{\bf P}^2+A^2=4$. These
operators act in 10-D spin space, and the kinematical restrictions
reduce this dimension to 7. Similarly to $SO(4)$ group, the vector
operators describe spin S=1 and transitions between spin triplet
and two singlet components of the multiplet, whereas the scalar
$A$ stands for transitions between two singlet states. Then the
vectors $\bf{R}$ and $\bf{P}$ may be identified with $\bf{R}_1$
and $\bf{R}_2$, and the spin algebra is connected with the algebra
for Hubbard operators by equations (\ref{SE}).

The group $SO(5)$ is non-compact, and the parametrization
(\ref{s051}) is not unique. As an alternative, one may refer to
another representation of $D_{\alpha\beta}$ used, the theory of
high-T$_c$ superconductivity \cite{zhang}:
\begin{eqnarray} -i\left(
\begin{array}{ccccc}
  0& &  & &  \\
  \pi_x+\pi_x^\dagger  & 0&  &  \\
  \pi_y+\pi_y^\dagger& -S^z& 0&  &  \\
  \pi_z+\pi_z^\dagger&S^y & -S^x& 0 & \\
  Q& -i(\pi_x^\dagger-\pi_x)& -i(\pi_y^\dagger-\pi_y)& -i(\pi_z^\dagger-\pi_z)&0
\end{array}
\right)\label{s052}
\end{eqnarray}
with 10 generators identified as a scalar $Q$ and three vectors
$\vec{S},\vec{\pi}$ and $\vec{\pi}^\dagger$ standing for the total
charge, spin and $\pi$ triplet $S=1$ superconductor order
parameter. Both representations (\ref{s051}) and (\ref{s052}) are
connected by the unitary transformation.

It is well known that the spin space may be projected onto bosonic
or fermionic  space by various transformations useful for specific
physical applications. Here we show the way of generalization of
several popular spin-boson and spin-fermion representations to the
dynamical algebras $o(4)$ and $o(5)$. Like in the case of pure
spin operators, these representations should preserve all
kinematical constraints.

Schwinger \cite{schw} has shown that it is possible to make a
realization of
 $SU(2)$ group for spin $S$ in terms of boson operators $a_i, a_i^\dagger$,
 $(i=1...2S+1)$ with
 $$[a_i,a_j^\dagger]=\delta_{ij}.$$
In particular, for $S=1/2$ Schwinger's bosonization reads
\begin{equation}
\vec{s}=\frac{1}{2} a^\dagger \vec{\tau} a, \;\;\;\;\;
a^\dagger=(a_1,a_2).
\end{equation}
($\vec{\tau}$ is the set of 2$\times$2 Pauli matrices). One should
introduce the local constraint $a_1^\dagger a_1+ a^\dagger_2 a_2
=1$  to get rid of spurious states associated with this bosonic
representation.

Using local isomorphism of $SO(4)$ and $SU(2)\times SU(2)$, one
may derive a representation of $SO(4)$ generators in terms of
4-component Schwinger bosons as
\begin{equation}
\vec{S}=\vec{s}_1+\vec{s}_2=\frac{1}{2}(a^\dagger \vec{\tau}_a a
+b^\dagger \vec{\tau}_b b),\;\;\;\;\;\;
\vec{R}=\vec{s}_1-\vec{s}_2=\frac{1}{2}(a^\dagger \vec{\tau}_a a
-b^\dagger \vec{\tau}_b b)
\end{equation}
where $4\times 4$ $\tau$-matrices act on $a-$ and $b-$ subsets of
4-vector ${\bf q}^T=(a_1^\dagger, a_2^\dagger, b_1^\dagger,
b_2^\dagger)$.

An alternative approach is based on use of Abrikosov \cite{Abrik}
or Popov-Fedotov \cite{pop, kis01} auxiliary  fermions
$f_\lambda$, where $\lambda=-1,0,1,s$. Making use of (\ref{m.1})
we represent 6 generators of $SO(4)$ as follows
\begin{eqnarray}
S^+ &=&  \sqrt{2}(f_0^\dagger f_{-1}+f^\dagger_{1}f_0),\;\;\;\;
S^- =  \sqrt{2}(f^\dagger_{-1}f_0+ f_0^\dagger f_{1}) ,\;\;\;\;
S^z =
f^\dagger_{1}f_{1}-f^\dagger_{-1}f_{-1},\nonumber \\
R^+  &=& \sqrt{2}(f^\dagger_{1} f_s -  f_s^\dagger f_{-1}),
\;\;\;\; R^- = \sqrt{2}(f_s^\dagger f_{1} - f^\dagger_{-1}f_s)
,\;\;\;\; R^z = -( f_0^\dagger f_s + f_s^\dagger f_0).
\label{4}\nonumber
\end{eqnarray}
with the only constraint
\begin{equation}
n_1+n_0+n_{-1}+n_s=1~,  \label{const}
\end{equation}
whereas the orthogonality condition is fulfilled automatically.
The constraint (\ref{const}) is respected by means of introducing
real chemical potential $\lambda \to \infty$ for Abrikosov's
auxiliary fermions or imaginary chemical potentials $\mu_t= -i\pi
T/3$ for Popov-Fedotov semi-fermions (see details in
\cite{kis00}). The advantage of semi-fermionic representation is
that it allows to construct a real-time Schwinger-Keldysh
formalism \cite{kis00} necessary for description of strongly
non-equilibrium effects in systems with dynamical symmetries.

The fermionic representation of $SO(5)$ group is easily
constructed by use of HO representation and is characterized by
5-vector ${\bf q}^T=(f_{-1}^\dagger f_0^\dagger, f_1^\dagger,
f_s^\dagger, f_r^\dagger)$
\begin{eqnarray}
S^+ &=&  \sqrt{2}(f_0^\dagger f_{-1}+f^\dagger_{1}f_0),\;\;\;\;
S^z =
f^\dagger_{1}f_{1}-f^\dagger_{-1}f_{-1},\nonumber \\
R^+  &=& \sqrt{2}(f^\dagger_{1} f_s -  f_s^\dagger f_{-1}),
\;\;\;\; R^z = -( f_0^\dagger f_s + f_s^\dagger f_0),\nonumber\\
P^+  &=& \sqrt{2}(f^\dagger_{1} f_r -  f_r^\dagger f_{-1}),
\;\;\;\;  P^z = -( f_0^\dagger f_r + f_r^\dagger f_0). \label{fff}
\end{eqnarray}
and
$$
A=i(f_r^\dagger f_s - f_s^\dagger f_r)
$$
The constraint
\begin{equation}
n_1+n_0+n_{-1}+n_s+n_r=1 \label{const2}
\end{equation}
is respected either by real infinite chemical potential (Abrikosov
pseudofermions) or by set of complex chemical potentials
(semi-fermions). We do not present here ten $5\times 5$ matrices
characterizing $SO(5)$ representation to save a space. The reader
can easily construct them using representations (\ref{SE}) or
(\ref{fff}). There exists also a bosonic representation based on
Schwinger bosons which might be derived by the method similar to
used above for $SO(4)$ group.

The representations of higher $SO(n)$ groups can be constructed in
a similar fashion. We address the reader to the papers \cite{KKA}
for further details of systematic classifications of $SO(n)$
groups where numerous examples of application of higher dynamical
groups for quantum dots can also be found.

Kinematic constraints imposed on auxiliary fermions and bosons is
in strict compliance with the Casimir and orthogonality
constraints in spin space. Accordingly, the number of fermionic
and bosonic fields reproduces the dimensionality of spin space
reduced by these constraints. We have seen that the 6-D space of
generators of $SO(4)$ group is reduced to D=4. Then the minimal
(unconstraint) fermionic representation for this group should
contain two $U(1)$ fermions. This means that the representation
(\ref{4}) is not minimal. Apparently, the best way to find such
representation is to use Jordan-Wigner-like transformation
\cite{Tsvel}. The same kind of arguments applied to $SO(5)$ group
tells us that the spinor field should contain seven components.
This means that the representation (\ref{fff}) should be completed
by one more (Majorana) fermion, and this fact points to one more
hidden $Z_2$ symmetry \cite{Tsvel}.
\section{Conclusion}
The concept of dynamical symmetry in quantum mechanics has been
(probably) first noted in the Hydrogen atom and the isotropic
Harmonic oscillator. In these systems, there is a high degree of
degeneracy of energy levels, which cannot be explained merely on
the basis of rotational symmetry of the relevant Hamiltonian. In
50-es and 60-es the approach based on the ideas of dynamical
symmetry was a powerful tool in high energy physics in an extremal
situation, when the theoreticians had no relevant Lagrangian for
description of experimentally observed hadron multiplets, but the
symmetry of the system could be restored by group-theoretical
methods, namely by constructing corresponding dynamical algebras.

Today, the concept of dynamical symmetry is ubiquitous in many
branches of modern physics, such as quantum field theory, nuclear
physics, quantum optics and condensed matter physics in low
dimensions. Quantum dots are especially suitable objects for the
group theoretical approach because the fully discrete spectrum of
low-lying excitations in these systems often may be characterized
by the definite dynamical symmetry, and the interaction with the
metallic reservoir of metallic electrons in the leads provides a
powerful tool of symmetry breaking.

In this review we concentrated  on the spin excitations in quantum
dots. Another promising class of nanoobject for applications of
these ideas is spin ladders. In this case the role of object with
definite dynamical symmetry is played by a single rung or pair of
neighboring rungs bound by diagonal bonds, whereas the
longitudinal modes violate this symmetry. The application of of
dynamical symmetry approaches in this field are seldom enough as
yet \cite{bled,bort}}, but the field seems to be really wide.

One more field for application of ideas developed in this review
is the rapidly developing area of molecular electronics
\cite{molec}. In artificially fabricated 2D molecular electronic
circuits individual organic molecules are incorporated in
electronic devises as basic elements.  Molecular bridges
containing one or several molecular groups and bound by tunnel
contact with the rest network are the potential objects for
description in terms of dynamical symmetry. Especially interesting
is the case when the magnetic ions are caged within such a
molecular group. One may expect manifestations of Kondo effect in
current-voltage characteristics of such objects and indeed, first
observations of Kondo resonances in tunnelling through
metallorganic molecules are available \cite{comol}. In such
objects the continuous symmetry of spin groups should be combined
with discrete symmetry of finite rotation groups, which
characterize the molecular symmetry. Possible involvement of
vibrational degrees of freedom open new horizons for studies of
dynamical symmetry at finite frequencies.

At the end, we hope to convince the reader of the beauty and
relevance of dynamical symmetries in condensed matter physics and
to stress its relation with down to earth experiments which,
following the impressive technological fabrication techniques, can
be performed in numerous laboratories.
\section*{ACKNOWLEDGMENTS}
We are grateful to L.W.Molenkamp and M.Heiblum for discussion of
various experimental aspects of nanophysics.  This work is
supported by the DFG under SFB-410 project, ISF grant, A.Einstein
Minerva Center and the Transnational Access Program $\#$
RITA-CT-2003-506095. MNK is grateful to Argonne National
Laboratory for the hospitality during his visit. Research in
Argonne was supported by U.S. DOE, Office of Science, under
Contract No. W-31-109-ENG-39.

\end{document}